\documentclass{aa}  

\usepackage{graphicx}
\usepackage{txfonts}
\usepackage{multirow}
\usepackage{xcolor,colortbl}
\usepackage{subcaption}
\begin{document}

%   \title{Protoclusters core SFR in hydrodynamical %cosmological simulations}
   \title{The DIANOGA simulations of galaxy clusters: characterising star formation in proto-clusters}

   %\subtitle{I. Overviewing the $\kappa$-mechanism}

   \author{L.\ Bassini\inst{1,2,3},
            E.\ Rasia\inst{2,3}, 
            S.\ Borgani\inst{1,2,3,4}, 
            G.L.\ Granato\inst{2,3,5}, 
            C.\ Ragone-Figueroa\inst{5,2},
            V.\ Biffi\inst{6,3},
            A.\ Ragagnin\inst{2}, 
            K.\ Dolag\inst{6}, 
            W.\ Lin\inst{7},
            G.\ Murante\inst{2}, 
            N.R.\ Napolitano\inst{7},
            G.\ Taffoni\inst{2},
            L.\ Tornatore\inst{2,3},
            Y.\ Wang\inst{7}
%          \fnmsep\thanks{Just to show the usage
%          of the elements in the author field}
          }

\institute{
Astronomy Unit, Department of Physics, University of Trieste, via Tiepolo 11, I-34131 Trieste, Italy
\and
INAF - Osservatorio Astronomico di Trieste, via Tiepolo 11, I-34131 Trieste, Italy\\
             \email{luigi.bassini@inaf.it}
\and
IFPU - Institute for Fundamental Physics of the Universe, Via Beirut 2, 34014 Trieste, Italy
\and
INFN - National Institute for Nuclear Physics, Via Valerio 2, I-34127 Trieste, Italy
\and
Instituto de Astronom\'ia Te\'orica y Experimental (IATE), Consejo Nacional de Investigaciones Cient\'ificas y T\'ecnicas de la\\ Rep\'ublica Argentina (CONICET), Universidad Nacional de C\'ordoba, Laprida 854, X5000BGR, C\'ordoba, Argentina
\and
Universitäts-Sternwarte München, Fakultät für Physik, LMU Munich, Scheinerstr. 1, 81679 München, Germany
\and
School of Physics and Astronomy, Sun Yat-sen University, Zhuhai Campus, 2 Daxue Road, Xiangzhou District, Zhuhai 519082, China
\\
}

\date{Received XXX; accepted XXX}
\titlerunning{Star formation in proto-clusters}
\authorrunning{Bassini et al.}

% \abstract{}{}{}{}{} 
% 5 {} token are mandatory
 
  \abstract
   {}
   {We studied the star formation rate (SFR) in cosmological hydrodynamical simulations of galaxy (proto-)clusters in the redshift range $0<z<4$, comparing them to recent observational studies; we also investigated the effect of varying the parameters of the star formation model on galaxy properties such as SFR, star-formation efficiency, and gas fraction.
   %We investigate the capability of state of the art numerical simulations to reproduce the high SFR observed in high redshift and highly star-forming protoclusters. We also test the possibility of reproducing galaxy properties (SFR, SFE, gas fraction) within this environment.
   }
   {We analyse a set of zoom-in cosmological hydrodynamical simulations centred on 
   twelve clusters. %within the mass range  $[1, 15]\times 10^{14}\ M_{\odot}$.
  %7 massive clusters ($M_{200} > 8 \times 10^{14}\ h^{-1} M_{\odot}$) and 5 smaller clusters ($M_{200} \in [1-4]\times 10^{14}\ h^{-1} M_{\odot}$). 
   %These simulations are a subsample of the Dianoga set, which comprises 29 clusters selected from a parent dark matter only simulation of $1\ h^{-1}\ \rm Gpc$ side. 
   The simulations are carried out with the GADGET-3 TreePM/SPH code which includes various subgrid models to treat unresolved baryonic physics, including AGN feedback. }
   {Simulations do not reproduce the high values of SFR observed within protoclusters cores, where the values of SFR are underpredicted by a factor $\gtrsim 4$ both at $z\sim2$ and $z\sim 4$. The difference arises as simulations are unable to reproduce the observed starburst population and is worsened at $z\sim 2$ because simulations underpredict the normalisation of the main sequence of star forming galaxies (i.e., the correlation between stellar mass and SFR) by a factor of $\sim 3$. As the low normalisation of the main sequence seems to be driven by an underestimated gas fraction, it remains unclear whether numerical simulations miss starburst galaxies due to a too low predicted gas fractions or too low star formation efficiencies. Our results are stable against varying several parameters of the star formation subgrid model and do not depend on the details of the AGN feedback.
   %starburst galaxies in protocluster enviroment are mainly driven by a high gas fraction or a high star formation efficiency, as observational results strongly depend on the assumptions needed to derive gas related quantities. Our results are stable within a wide range of values for the parameters of the star formation subgrid model and does not depend on the implementation of AGN feedback.
   }
   {The subgrid model for star formation (\citealt{2003SPRINGEL}), introduced to reproduce the self-regulated evolution of quiescent galaxies, is not suitable to describe violent events like high-redshift starbursts. We find that this conclusion holds independently of the parameters choice of the star formation and AGN models. The increasing amount of multi-wavelength high-redshift observations will help improving the current star formation model, in order to fully recover the observed star formation history of galaxy clusters.

   %The results indicate that the model of star formation should be modified in order to match the observed star formation history of galaxy clusters. Our results highlight that observations of strong star formation in proto-cluster regions, combined with observations of the old stellar populations in low-redshift cluster galaxies, provide non-trivial constraints on models of galaxy evolution in the extreme (proto-)cluster environment. 
   
   %The increasing number of multi-wavelenght high-redshift observations will help to improve the model.
   %A better modelling, based on a deeper understanding of the physics involved, is needed to match the observational constraints that are becoming available at $z>2$.
   }
   %and give to simulations the credibility to make reliable predictions.}

\maketitle
%
%-------------------------------------------------------------------

\section{Introduction}
%https://arxiv.org/pdf/1501.04105.pdf
%https://arxiv.org/pdf/1911.09368.pdf
    
    Galaxy clusters are the most massive objects in the universe; they are located at the nodes of the cosmic web and are characterised by very high densities. Given their extreme nature, they are very important probes for both cosmology (\citealt{2012KRAVTSOV}) and galaxy formation models, where the evolution of galaxies and thus their resulting properties depend on a range of environmental effects. 

    In the local universe, galaxy clusters appear as massive virialized objects. They are characterised by the presence of a hot ($\sim 10^8\ \rm K$) diffuse plasma, the intracluster medium (ICM), which can be studied through X-ray observations and the Sunyaev-Zeldovich effect (\citealt{2002ROSATI}, \citealt{2002CARLSTROM}). Galaxies, whose velocity dispersion is consistent with the ICM temperature, are mainly bulge dominated and ellipticals with reduced ongoing star formation, especially in the cluster core region. The stellar population is typically very old, with stars nearly as old as the Universe. Indeed observations (\citealt{2010MANCONE}, \citealt{2014WYLEZALEK}, \citealt{2015FOLTZ}) and theoretical semi-analytical models (e.g., \citealt{2007DELUCIA}) suggest that galaxies are undergoing passive evolution since $z\sim 1$ and have formation times $z_f \gtrsim 2$. 

    Differently from the local universe where all galaxy clusters show such similar properties, at $z \gtrsim 1.4$ they behave as a rather diverse population. Some observations report the discovery of already mature clusters, with an enhanced fraction of red and quenched galaxies with respect to the field, at least in the core (\citealt{2010PAPOVICH}, \citealt{2010STRAZZULLO}, \citealt{2011GOBAT}, \citealt{2013STRAZZULLO}, \citealt{2013TANAKAb}, \citealt{2014NEWMAN}, \citealt{2014ANDREON}, \citealt{2016COOKE}), while other clearly show a mixed population of quenched and star forming galaxies (\citealt{2013TANAKA}, \citealt{2013BRODWIN}, \citealt{2013GOBAT}, \citealt{2017HATCH}, \citealt{2016STRAZZULLO}). Few observations also suggest a revers star formation rate (SFR)-density relation at $z\gtrsim 1.5$, where the specific star formation (sSFR) increases toward the core of the cluster (\citealt{2010TRAN}, \citealt{2014SANTOS}, \citealt{2015SANTOS}, \citealt{2019SMITH}).

    At even higher redshift, $z \gtrsim 2$, systems lack the presence of a massive virialized halo showing, instead, multiple halos spread over large scales (\citealt{2012HAYASHI}, \citealt{2014LEMAUX}, \citealt{2015KUBO}, \citealt{2015CASEY}, \citealt{2016CASEY}). This is in line with theoretical expectations from numerical simulations, which predict a hierarchical formation of massive clusters formed by the assembly of smaller halos that at $z \sim 2$ might occupy a region as large as $20\ \rm comoving\ Mpc$ (cMpc). (\citealt{2013CHIANG}, \citealt{2015MULDREW}, \citealt{2016CONTINI}).

    Since at this redshift (proto)clusters are not virialized, it is difficult to detect them with techniques based on the ICM properties. Thus different methods have been used, like galaxy over-densities. This approach, however, can bias the results, depending on the galaxy properties used for the selection. In this respect, an important population of galaxies are dusty star forming galaxies (DSFG, see \citealt{2014CASEY}), highly star forming and heavily obscured by dust, emitting in the far infrared (FIR) and sub-millimetric bands. These galaxies represent the strongest starbursts and are expected to be the progenitors of local massive ellipticals (\citealt{2008CIMATTI}, \citealt{2010RICCIARDELLI}, \citealt{2013FU}, \citealt{2013IVISON}, \citealt{2014TOFT}, \citealt{2018GOMEZ}). They trace the dusty star-forming phase of protoclusters, and their expected short star-burst phase of few hundreds of Myrs (\citealt{2004GRANATO}) makes them relatively rare objects in the sky. Albeit their rareness they have been successfully used to identify dense and highly star forming environments up to redshift $z\sim 4$ (\citealt{2014CLEMENTS}, \citealt{2018OTEO}, \citealt{2018MILLER}) and have been observed in several already known high redshift protoclusters (\citealt{2009CHAPMAN}, \citealt{2014DANNERBAUER}, \citealt{2015UMEHATA}, \citealt{2018COOGAN}, \citealt{2019LACAILLE}, \citealt{2019SMITH}). 
    
    %and a peak of star formation rate density (SFRD) for (proto)clusters at $2 \lesssim z \lesssim 3$, roughly in agreement with the peak of the cosmic SFRD (\citealt{2014MADAU}, \citealt{2019BOCO}). Along with observations, also semi-analytical models suggest that most of the stars that end up in clusters are formed at high redshift (\citealt{2007DELUCIA}). Moreover, if in the local universe there is a positive correlation between high densities and fraction of quiescient glaxies, (proto)cluster regions at $z\sim 2$ are expected to contribute to $\sim 20\%$ of the cosmic SFRD budget (\citealt{2017CHIANG}) with observations suggesting a reversal in the SFR-density relation at $z\gtrsim 1.5$ (\citealt{2010TRAN}, \citealt{2014SANTOS}, \citealt{2015SANTOS}, \citealt{2019SMITH}).
    
    So far, numerical simulations have been unable to reproduce the high SFRs observed in protoclusters characterised by overdensities of DSFGs (\citealt{2015GRANATO}), as simulations miss to predict sufficiently high peaks of star formation activity at early epochs. This result adds to a long standing difficulty for cosmological simulations to reproduce star formation properties of galaxies, such as the main sequence of star forming galaxies, around the peak of the cosmic star formation rate density (\citealt{2016DAVE}, \citealt{2017MCCARTHY}, \citealt{2019DONNARI}, \citealt{2019DAVE}). Indeed, \cite{2015GRANATO} found that the bulk of star formation within the observed putative progenitors of massive galaxy clusters occurred at higher rates and lasted less than in simulations. This conclusion was based on the observations available at that time characterised by low angular resolution and SFR integrated on the Mpc scale. In the last few years, with instruments like ALMA, it has been possible both to resolve single sources within protoclusters and have information on the galaxy cold gas content (\citealt{2018WANG}, \citealt{2019GOMEZ}, \citealt{2020HILL}). On the simulations side, progress has been made to increase the numerical resolution, needed to resolve higher density peaks and related higher SFRs. Therefore, times are ready for a deeper inspection on the simulations capability of reproducing protocluster properties.
    
    In this work we make use of $12$ simulations out of the set of $29$ hydrodynamical zoom-in simulations of galaxy clusters named Dianoga, to investigate the predictive power of state of the art cosmological simulations around the peak of the SFR in the protocluster stage of structure formation. In particular, we aim at comparing both the integrated values of SFRs in protocluster regions and the protocluster galaxies properties to recent observations that are now available. The set of simulations used is particularly suited for this aim, as it comprises massive objects that can only be found in a fair number within large, $\sim 1\ h^{-1}\ \rm Gpc$ a side, cosmological boxes. Moreover, the simulations has been carried out at a resolution 10 times higher than before (i.e, \citealt{2015GRANATO}). 

    The paper is structured as follows: in Sect.~2 we describe the simulations set up, with particular focus on the AGN feedback implementation and the subgrid star formation model. In Sect.~3 we show brightest cluster galaxies (BCGs) properties and stellar mass function at $z=0$. In Sect.~4 and Sect.~5 we compare the predicted SFRs in protocluster regions with the available observations at $z\sim 2$ and $z\sim 4$ respectively, and we analyse the main sequence of star forming galaxies at both redshifts. In Sect.~6 we show the evolution of the mass normalised SFR in clusters and protoclusters. In Sect.~7 we study gas related properties of our simulated galaxies, in comparison with observations. In  Sect. 8 we summarise the main results of our analysis and draw the main conclusions. 
    
%--------------------------------------------------------------------

%-------------------------------------------------------------------

\section{Simulations}

    In this section we describe the set of numerical simulations used in this work, and in particular we detail the observational constraints used to calibrate the subgrid model of AGN feedback. We also briefly review the star formation model of \cite{2003SPRINGEL} implemented in our code.
    
    \subsection{Set-up of simulations}
    The analysis presented in this paper is based on a set of 12 hydrodynamical zoom-in simulations evolved in a $\Lambda$CDM cosmology, with parameters: $\Omega_{\rm m} = 0.24$, $\Omega_{\rm b}=0.037$, $n_{\rm s}=0.96$, $\sigma_8=0.8$ and $H_0 = 100h\ \rm km\ s^{-1}\ Mpc^{-1} =72\ \rm km\ s^{-1}\ Mpc^{-1}$. These are part of a sample of 29 simulations, the Dianoga set, described in \cite{2015RASIA}, \cite{2017PLANELLES}, \cite{2017BIFFI}, \cite{2018BIFFI}, \cite{2018RAGONE}, and \cite{2019BASSINI}, but have a 10 times higher mass resolution and small differences in the code (see below) with respect to the cited works. We will refer to the previous set of simulations as low resolution (LR) simulations throughout the paper. The regions are extracted from a parent dark-matter (DM) only simulation of $1\ h^{-1}\ \rm Gpc $ side. From this cosmological box the 24 most massive clusters ($M_{200} > 8\times 10^{14}\ h^{-1} \rm M_{\odot}$)\footnote{We define $R_{\Delta}$ as the radius of the sphere encompassing an average density $\Delta$ times the critical density of the universe at that redshift, $\rho_{\rm crit}(z) = 3H^2(z)/8\pi G$. $M_{\Delta}$ will be the mass within $R_{\Delta}$.} were selected together with 5, randomly chosen, smaller objects ($M_{200}\in [1-4]\times 10^{14}\ h^{-1}\ \rm M_{\odot}$). Their Lagrangian regions, of radius about 5 times the virial radius of the selected clusters, were re-simulated with the inclusion of baryons and at a greater resolution (see \citealt{2011BONAFEDE} for a full description of the re-simulation procedure). The 12 simulations used for this work include the 5 less massive clusters and 7 massive clusters. The masses of the particles in the high resolution region are $m_{\rm DM} = 8.44\times 10^7\ h^{-1} \rm M_{\odot}$ for DM and $m_{\rm gas} = 1.56\times 10^7\ h^{-1} \rm M_{\odot}$ for the initial gas particles. The Plummer equivalent gravitational softening adopted for DM particles is $4.2\ h^{-1}$ comoving kpc (ckpc) at $z>2$ and $1.4\ h^{-1}$ physical kpc (pkpc) otherwise. The softening lengths for gas, star, and black holes (BHs) particles are $1.4$, $0.35$, and $0.35\ h^{-1}$ pkpc respectively.
    
    The simulations are carried out with the code GADGET-3, a modified version of the Tree-PM Smoothed-Particle Hydrodynamics (SPH) public code GADGET2
    (\citealt{2005SPRINGEL}). We employed the hydrodynamical scheme presented in \cite{2016BECK}, where a higher order interpolating kernel function is implemented, along with a time-dependent artificial viscosity and a time-dependent artificial conduction, which improve the SPH performance in capturing discontinuities and the development of gas-dynamical instabilities. The unresolved baryonic physics, mostly involving the stellar component and the activity of the BH population, is treated with sub-grid models. The prescription of metal-dependent radiative cooling follows \cite{2009WIERSMA}. The model of star formation and associated feedback is implemented according to the original model by \cite{2003SPRINGEL}, see below for extra details. For the metal enrichment and chemical evolution we follow the formulation by \cite{2007TORNATORE}. The stellar yields and a deeper description of metals are specified in \cite{2018BIFFI} (see also, \citealt{2017PLANELLES} and \citealt{2017BIFFI}). The treatment of BH and associated AGN are detailed below (Sect.~\ref{sec_agn_intro}) after a brief summary of the star formation model.

    %A detailed description can be found in \cite{2017PLANELLES} or \cite{2017BIFFI} (see below for relevant differences in the treatment of the BH activity); we briefly summarize here the main aspects. The prescription of metal-dependent radiative cooling follows \cite{2009WIERSMA}. The model of star formation and associated feedback is implemented according to the original model by \cite{2003SPRINGEL}, while for the metal enrichment and chemical evolution we follow the formulation by \cite{2007TORNATORE}. The stellar yields used in our simulations are specified in \cite{2018BIFFI}. In the following we detail the AGN feedback and star formation models.
    
        \subsection{Star formation}\label{sec:effective_model}
    
    Here we review the main aspects of the subgrid model for the star formation, which we will discuss throughout the paper. For a full explanation we refer to the original paper by \cite{2003SPRINGEL}.
    In this model each SPH particle samples a region of the interstellar medium (ISM), and is subdivided in a cold phase and a hot phase characterised by densities  $\rho_c$ and $\rho_h$, in pressure equilibrium one with each other. The total density associated to a particle will be the sum of the two: $\rho = \rho_c + \rho_h$. A SPH particle has a non null fraction of cold gas (i.e., $\rho_c>0$), and thus becomes multiphase, whenever its density is higher than a given threshold $\rho_{\rm thr}$. Given the cold fraction, a numerical instantaneous SFR is associated to each multiphase particle\footnote{By numerical SFR we mean the rate at which the mass in SPH gas particles should be transformed into stellar particles. The actual physical SFR of the model is $\rho_c/t_{\star}$}:
    \begin{equation}\label{eq:em1}
        \frac{{\rm d} \rho_{\star}}{{\rm d}t} = \dot{\rho}_{\star} = 
        \left( 1 - \beta \right) \frac{\rho_c}{t_{\star}},
    \end{equation}
    
    \noindent where $t_{\star}$ is the characteristic timescale for star formation, while $\beta$ is the fraction of massive stars that are expected to instantly explode as supernovae and depends on the chosen IMF. In this work, we employ a Chabrier initial mass function (IMF, \citealt{2003CHABRIER}). The parameter $t_{\star}$ follows the expression:
    \begin{equation}\label{eq:em2}
        t_{\star}(\rho) = t^{\star}_0 \left( \frac{\rho}{\rho_{\rm thr}} \right)^{-1/2}.
    \end{equation}
    
    \noindent $t^{\star}_0$ is set to $1.5\ \rm Gyr$ in order to match the observed Kennicutt relation (\citealt{1998KENNICUTT}). In practice, varying this parameter directly reflects into a variation of the numerical star formation efficiency (SFE). Indeed, using eq.~\ref{eq:em1} and eq.~\ref{eq:em2}:
        
    \begin{equation}
        {\rm SFE} = \frac{\dot{\rho}_{\star}}{\rho_{c}} = \frac{1-\beta}{t_0^{\star}} \left( \frac{\rho}{\rho_{\rm thr}} \right)^{1/2}.
    \end{equation}
    
    \noindent Given the SFR, part of the cold clouds is evaporated by means of supernovae feedback:
    \begin{equation}\label{eq:em3}
        \frac{{\rm d}\rho_c}{{\rm d}t} = -A \beta \frac{\rho_c}{t_{\star}},
    \end{equation}
    
    \noindent where $A$ is the efficiency of evaporation that determines the efficiency of thermal supernovae feedback and is taken to be a function of the local gas density, $A \propto \rho^{-4/5}$. These equations give rise to the self-regulated cycle of star formation: high cold cloud density leads to a high SFR, that in turn means more feedback and cloud evaporation. When cloud evaporates, the SFR decreases and material is returned to the hot phase, increasing $\rho_h$. Finally, a higher density means a higher cooling rate, which causes more gas to condense in cold clouds, so that the cycle restarts. 
    
%    \subsection{AGN feedback}\label{AGN feedback}
    \subsection{AGN feedback}\label{sec_agn_intro}
    
    The AGN feedback is inspired to the original model developed by \cite{2005SH} and is implemented according to the scheme described in \cite{2013RAGONE} with two main modifications. First, in the feeding process we differentiate between hot and cold accretion (see Sect.~\ref{sec:AGN_fb}, Eq.~\ref{eq:bondi}). Second, we do not impose a temperature threshold to define multiphase gas particles and the energy released by AGN feedback is not used to evaporate the cold phase of gas particles. This second modification is motivated by the fact that this implementation results in a better agreement between the simulated galaxy stellar mass function (GSMF) and the observed one (see Sect.~\ref{sec:gsmf}), with the side effect of producing too massive BCGs at $z=0$ (see Sect.~\ref{sec:mstar-Mbcg}). 
    
    \subsubsection{BH seeding and positioning}\label{AGN_positioning}
    
    Briefly, during run time we identify groups of particles using the Friends of Friends (FoF) algorithm (\citealt{1982HUCHRA}). In practice, two DM particles are considered to be part of the same group if their distance is less than a fixed parameter, referred as linking length, which is commonly defined as a fraction of the mean inter-particle distance, $\bar{d}$. We fix this parameter to $0.16\times \bar{d}$. Hence, BHs in our simulations are spawned at the centre of each FoF group (defined as the position of the most bound particle) with a seed mass of $5.5\times 10^5\ \rm M_{\odot}$ whenever all these conditions are simultaneously fulfilled: $(i)$ the total stellar mass is higher than $2.8\times 10^9\ \rm M_{\odot}$; $(ii)$ the stellar to DM mass ratio is higher than $0.05$; $(iii)$ the gas mass is equal or larger than 10 percent of stellar mass; $(iv)$ no other central BH is already present.
    As the simulation evolves, we avoid the presence of wandering BHs by adopting a different strategy from \cite{2018RAGONE}. In this previous work we pinned the BHs, meaning that we re-positioned them at each time step at the location of the most bound particle of a halo. Here, instead, we assign to the BH a large dynamical mass and we use low values of star and BH particles softening lengths. Namely, the BH dynamical mass is imposed to be equal to the DM particle mass until it outgrows that value and the softening values are four times smaller than before, once re-scaled to the higher resolution. These numerical prescriptions are sufficient to mimic a dynamical friction without the necessity to explicitly include a dynamical friction force (\citealt{2016STEINBORN}). Even though this scheme performs overall well at the current numerical resolution, BH centring remains a major challenge for our, and presumably all, numerical simulations, and it still can happen that a BH moves from the centre of a structure. This is particularly problematic in cluster simulations, where the absence of AGN feedback at the centre of the BCG would lead to catastrophic cooling, with resulting high BCG mass and SFR of the order of $\sim 10^{3}\ \rm M_{\odot}\ \rm yr^{-1}$ at $z\sim 0$. Indeed, in 1 out of the 12 simulations used for this work, the proto-BCG looses its central BH at $z\sim 4$, and is characterised by an incredibly high SFR at $z=0$. Even though it is not an issue for the conclusions of this work (see Sect.~\ref{sec:discussion}), the problem needs to be addressed in future simulations.

    %Differently from the prescription employed in \cite{2018RAGONE} and in the other previous works, the BH particle is not repositioned at the position of the most bonded particle at each time step. Instead, we avoid wandering BHs employing a dynamical mass for the BH particle equal to $m_{\rm DM}$ until $M_{\rm BH} > m_{\rm DM}$ (after that the dynamical mass equals the actual mass of the BH particle) and reducing the BH and stellar particles softenings by a further factor of $2$ with respect to the scalings due to the increased mass resolution.
    \subsubsection{AGN accretion and feedback}\label{sec:AGN_fb}
    
    Once BHs are seeded, they grow by two different channels: accretion of the surrounding gas and BH-BH mergers. The former follows the Eddington-limited alpha-enhanced Bondi accretion rate (\citealt{1952BONDI}) formula:
    \begin{equation}
        \dot{M}_{\rm Bondi, \alpha} = \alpha \frac{4 \pi G^2 M^2_{\rm BH} \rho}{(c_s^2 + v_{\rm BH}^2)^{3/2}},
        \label{eq:bondi}
    \end{equation}
    
    \noindent with $\alpha$ equal to 10 and 100 for hot ($T>5\times 10^5\ K$) and cold ($T<5\times 10^5\ K$) gas respectively (\citealt{2015STEINBORN}). In eq. \ref{eq:bondi} all gas-related quantities (sound speed, $c_s$, bulk gas velocity relative to BH velocity, $v_{\rm BH}$, and gas density, $\rho$) are smoothed over $200$ gas particles with a kernel function centred at the position of the BH. Given the gas accretion onto the BH particle, AGN energy feedback is given by
    
    \begin{equation}
        \dot{E} = \epsilon_r \epsilon_f \dot{M} c^2
        \label{eq:AGN_f}
    \end{equation}
    
    \noindent where $\dot{M}$ is the minimum between Eq.~\ref{eq:bondi} and the Eddington
    accretion rate, $\dot{M}={\rm min}(\dot{M}_{\rm Bondi, \alpha}, \dot{M}_{\rm Eddington})$, and the energy is distributed and thermally coupled to the nearest $200$ gas particles. In eq. \ref{eq:AGN_f}, $\epsilon_r$ is the fraction of mass transformed in radiation energy and 
    $\epsilon_f$ is the fraction of radiated energy thermally coupled to the gas particles.  
    %
    %ER: employed to determine the gas accretion through eq. \ref{eq:bondi}. 
    In the previous version of the code the energy was used to evaporate the cold fraction of the multiphase gas particles (see Sect.~\ref{sec:effective_model} for a brief explanation of multiphase particles; for a comprehensive review we remand to the original paper of \citealt{2003SPRINGEL}), while in the current set up we couple the energy only to the hot phase of each gas particle. The effects of this choice on our results are presented in Sect.~\ref{sec:tests}.
    
    BH particles can also grow by BH-BH mergers. In our implementation of the subgrid model two BHs are allowed to merge whenever all these conditions are fulfilled: $(i)$ $v_{\rm rel} < 0.5 \times c_s$; $(ii)$ $r_{\rm rel} < 3.5\ h^{-1}\ \rm pkpc$; $(iii)$ $|V_{\rm pot, rel}| + v_{\rm rel}^2 < 0.5\times c_s^2$; where $v_{\rm rel}$ and $r_{\rm rel}$ are the relative velocity and position between the two BH particles, $c_s$ is the sound speed and $V_{\rm pot, rel}$ is the difference between the gravitational potentials computed at the positions of the BH particles.

    %In \ref{fig:bcg_sfr} we compare SFR in BCGs with observations. In this case we select only BCGs at the center of clusters with $M_{2500} > 10^14M_{\odot}$. It appears that simulated BCGs have higher residual SFR than observed ones. However, BCGs residing in less massive clusters appear to have 0 SFR. The feeling is that the feedback used to have a 'low' SFR in massive halos kills sfr in lower massive systems. A comparison with other simulations (Eagle and IllustrisTNG) can be done using the data in https://arxiv.org/pdf/1908.11380.pdf. Here they compute the ssfr averagin over 300 Myrs. The aperture is not specified but I assume is the same used in other papers (https://arxiv.org/pdf/1906.02747.pdf): 2 times the halm mass radius. We can not use the same procedure but an aperture of 50 kpc should do the work.

    \begin{figure}
        \centering
        \includegraphics[width=\linewidth]{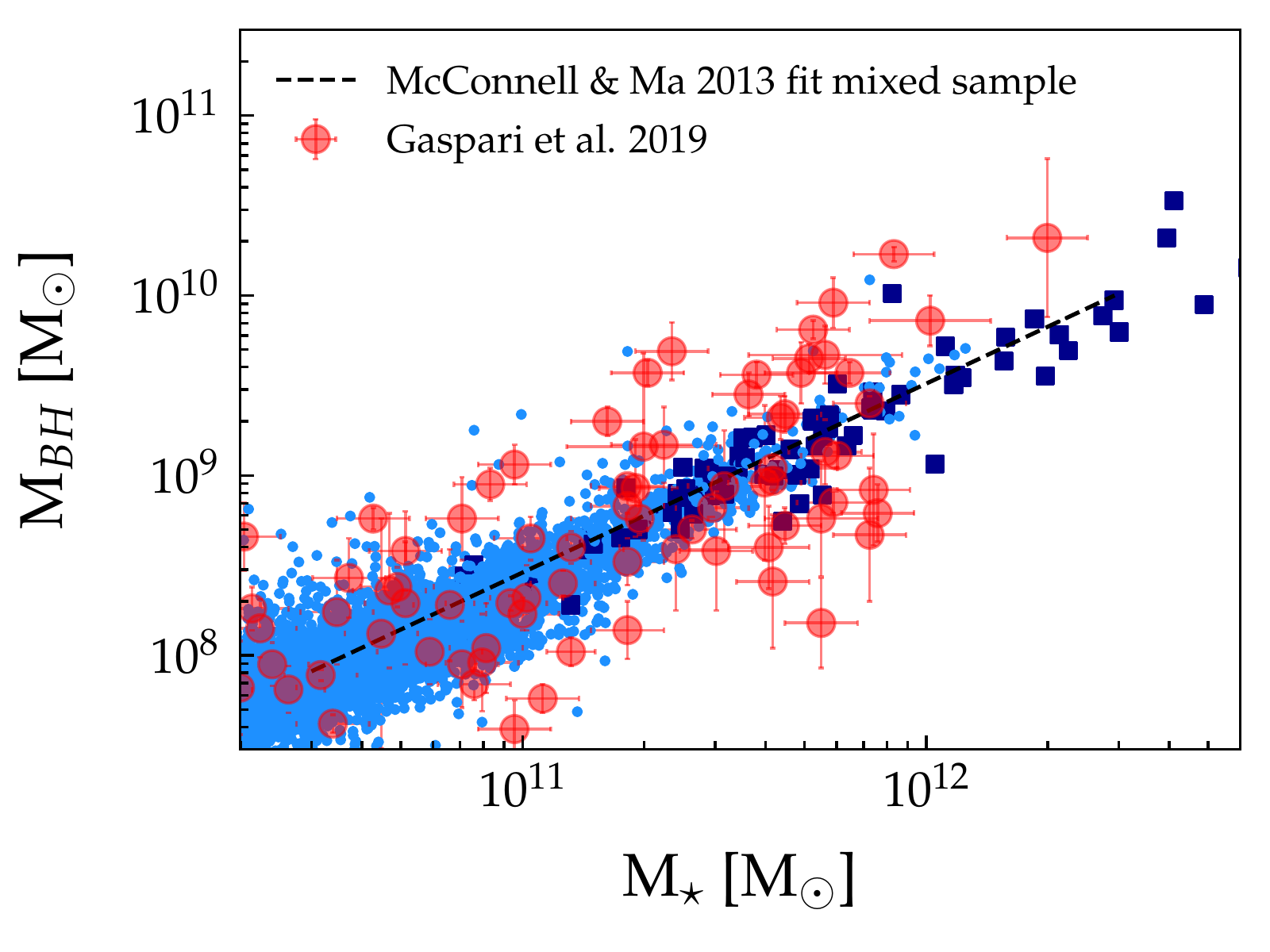}
        \caption{Correlation between the galaxies stellar mass and the central SMBHs mass. Observational data are taken from \cite{2013MCCONNELL} (dashed black line) and from \cite{2019GASPARI} (red circles). The simulated stellar masses for satellite galaxies (cyan points)  are obtained considering the star particles, bound to the substructure (accordingly to Subfind) and within 50 pkpc from its centre. The mass of the central galaxies (dark-blue squares) is obtained by summing over all stellar particles within an aperture of $0.15\times R_{500}$. }
        \label{fig:magorrian}
    \end{figure}

  \subsubsection{AGN feedback calibration} 
    The values of $\epsilon_f$ and $\epsilon_r$ are chosen in order to reproduce the observed 
    %normalization of the 
    correlation between BH mass and galaxy stellar mass (\citealt{1998MAGORRIAN}). In particular, we aim at reaching agreement with the observational results reported by \cite{2013MCCONNELL} and more recently by \cite{2019GASPARI}. The value chosen for $\epsilon_r$ is $0.07$ independently of  $\dot{M}_{\rm Bondi, \alpha}$, while $\epsilon_f$ is lower than $1$ only in quasar-mode, when $\dot{M}_{\rm Bondi,\alpha} / \dot{M}_{\rm Eddington}$ > 0.01 ($\epsilon_f = 0.15$). 
    In Fig.~\ref{fig:magorrian} we show numerical results in comparison with observations. In the plot dark-blue squares are simulated central galaxies, defined as galaxies at the centre of groups with at least $100$ substructures. All other simulated galaxies are represented as light-blue points. Stellar masses of non-central galaxies are computed considering the star particles associated to the substructures identified by the code Subfind (\citealt{2009DOLAG}) and within a sphere of $50$ pkpc. Stellar masses of central galaxies are computed considering an aperture of $0.15 \times R_{500}$ to match the aperture used by \cite{2019GASPARI}. We note that even though simulations correctly reproduce the normalisation of the observed correlation, the scatter is still under-reproduced, especially at the high mass end. Indeed, the intrinsic scatter around the observed correlation of \cite{2019GASPARI} is $\sigma = 0.40 \pm 0.03$, while it is a factor of $2$ lower in our simulations ($\sigma = 0.20$)\footnote{For the linear regression we used the public python package {\it linmix} (https://github.com/jmeyers314/linmix), which accounts for measurement errors in both the dependent and independent variables.}. Even though the observed scatter can be marginally boosted by uncertainties on the assumptions made to obtain these quantities from observational data (e.g., star formation history, IMF, metallicity, etc.), the most probable explanation of this difference is that the subgrid models adopted do not capture the diversity of conditions of BH accretion and AGN feedback at small scales.
  
    %In the most realistic case, the discrepancy is due to a combination of the two. 
    
%---------------------------------------------------

    \section{Galaxy cluster population at $z=0$}\label{sec:result_z0}
    In this section we analyse GSMF and BCGs properties at $z=0$ in our simulations in comparison with observations. 
    \begin{figure}
        \centering
        \includegraphics[width=\linewidth]{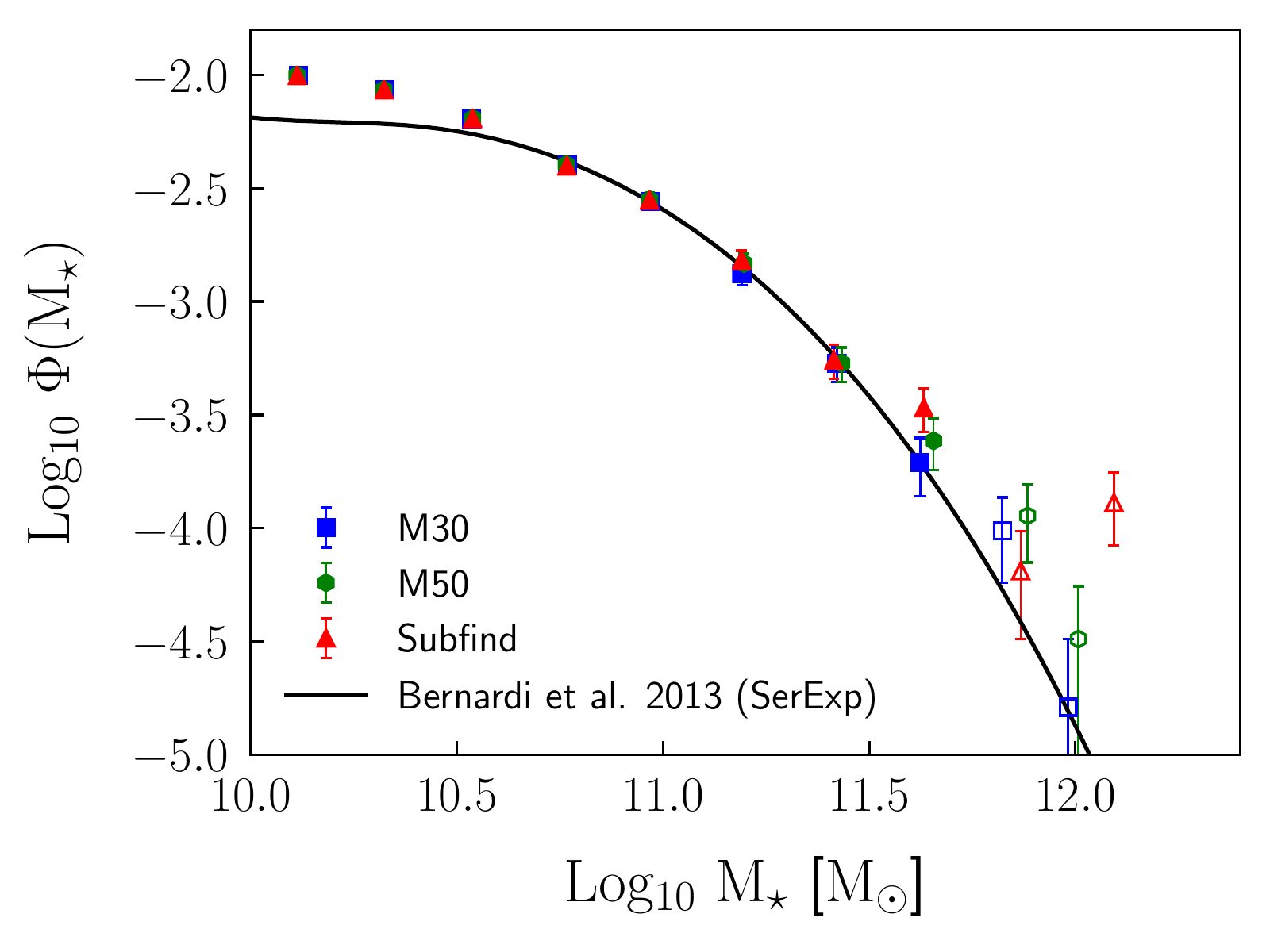}
        \caption{GSMF at $z=0$. Observational data are taken from \cite{2013BERNARDI} (black solid line). Simulations data are derived considering as stellar mass the sum of all stellar particles bound to the galaxy by Subfind (red triangles), and the same sum restricted to particles within 50 pkpc (green hexagon) and 30 pkpc (blue squares). Error bars are computed assuming Poissonian errors. The simulated GSMF is normalised following Eq.~\ref{eq:vulcani}. Filled and empty marks represent the mass bins with respectively more than and less than 10 galaxies. }
        \label{fig:smf}
    \end{figure}
    
    \subsection{Galaxy stellar mass function}\label{sec:gsmf}
    We start by comparing the observed and simulated GSMF. First we need to take into account the fact that our simulations are centred on galaxy clusters, while we compare to data obtained for the field. For this reason, we expect to have in simulations a significant higher normalisation of the GSMF. Operatively, we define galaxy clusters as spherical regions enclosing a mean matter density $\Delta$ times the critical density, $\rho_{\rm crit}$. Using the relation between $\rho_{\rm crit}$ and the mean cosmic matter density $\bar{\rho}$: 
    
    \begin{equation}
        \rho_{\rm crit}(z) = \bar{\rho}(z)\times [\Omega_{\rm M}(1+z)^3+\Omega_{\Lambda}] / \Omega_{\rm M}(1+z)^3,
    \end{equation}
    
    \noindent and assuming that galaxies follow the DM distribution, we have to normalise our GSMF by
    
    \begin{equation}\label{eq:vulcani}
        N_{\rm norm} = \Delta \times [\Omega_M(1+z)^3+ \Omega_{\Lambda}]/\Omega_M(1+z)^3,
    \end{equation}
    
    \noindent (see \citealt{2014VULCANI}, appendix B). In practice, we consider the most massive cluster in each of our simulated regions and all the galaxies but the BCG within $R_{100}$ (i.e., $\Delta=100$), and then we normalise each mass bin by $N_{\rm norm}$.

    The stellar mass of each galaxy is computed using three different definitions: the sum of all stellar particles that Subfind associates to a substructure and the same sum limited to stellar particles within 30 pkpc and 50 pkpc from the centre. The results are shown in Fig.~\ref{fig:smf}, where we plot the GSMF obtained with the three definitions of stellar mass. Results from simulations agree overall well with observations from \cite{2013BERNARDI} starting from the stellar mass of galaxies that we consider well-resolved in our simulated set ($M_{\star} > 10^{10}\ M_{\odot}$). The main difference is around $10^{10} \ \rm M_{\odot}$, were simulations have too many galaxies. As noted by \cite{2019HENDEN}, larger values of the softening length numerically decrease the number of galaxies at these masses. However, an investigation of the stellar feedback will be needed to better describe the GSMF at our low mass end. Regarding the different stellar mass definitions, there is no statistical difference in the results once a fixed aperture is used, as Subfind likely associate to massive galaxies also a fraction of the intracluster light (ICL). The good agreement with observational data is an improvement with respect to the GSMF presented in \cite{2019BASSINI} and obtained with the set of simulations at lower resolution. In that case we showed that the GSMF was a factor of $\sim 2$ below the results of \cite{2013BERNARDI} at $M_{\star} \sim 10^{11}\ M_{\odot}$. This difference, as we will discuss in the next subsection, is not due to the increased resolution, but to the different implementation of the AGN feedback discussed in Sect.~\ref{sec_agn_intro}.

    %On the contrary, as already noted in \cite{2019BASSINI}, the normalization of the GSMF in the LR runs is $\sim 3$ lower than observed. This result is mainly due to a higher SFR at higher redshift, resulting from the higher resolution and consequent possibility to resolve higher gas densities. 
    
    \subsection{Properties of the BCG}    
    
    \begin{figure}
        \centering
        \includegraphics[width=\linewidth]{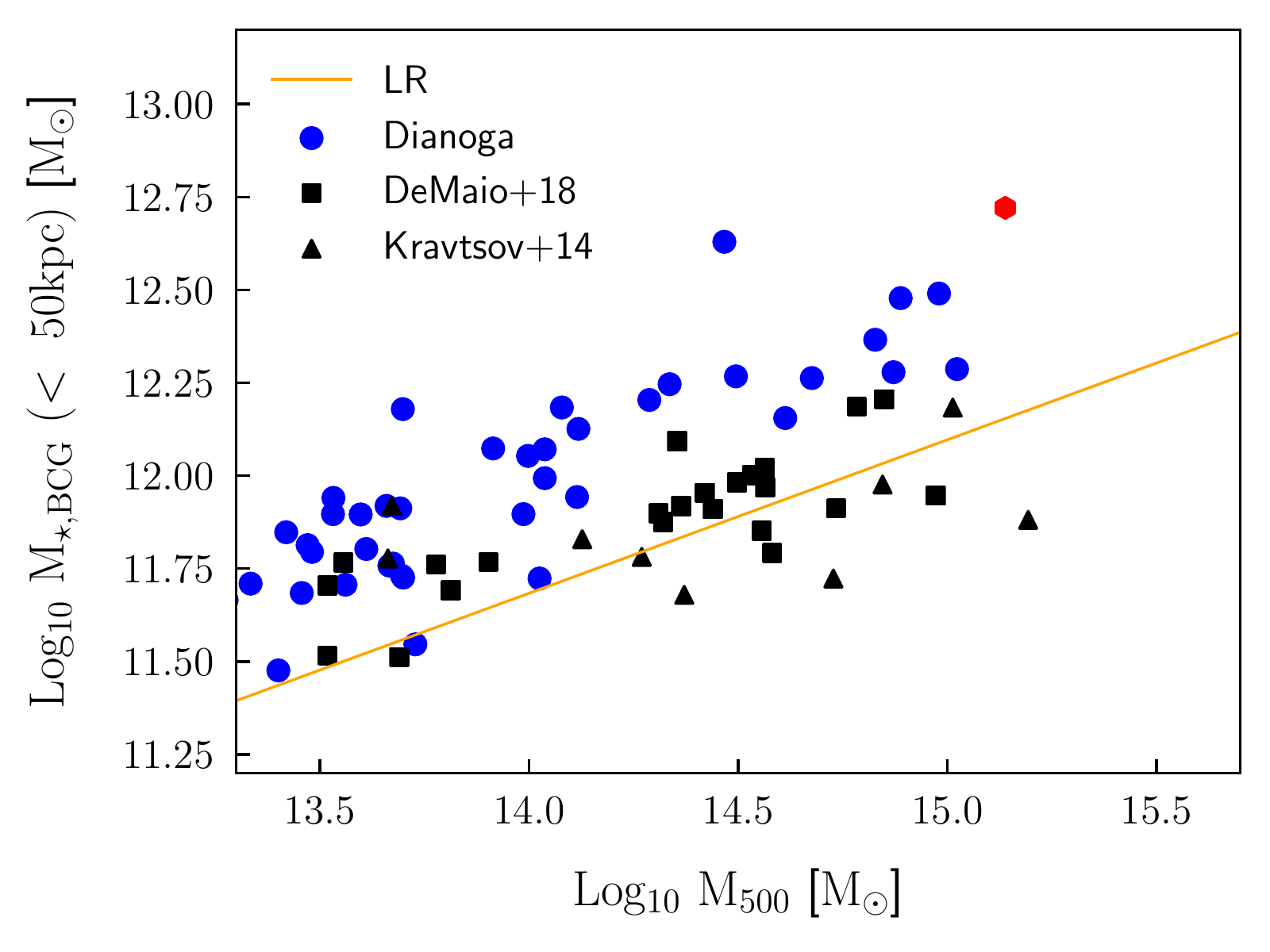}
        \caption{Correlation between BCGs stellar mass and $M_{500}$ at $z=0$. Observations are taken from \cite{2018DEMAIO} (blacks quares) and \cite{2018KRAVTSOV} (black triangles). The simulated values are shown as blue points. The red hexagon refers to the BCG that lost its central BH (see Sect.~\ref{AGN_positioning}). The orange line is the fit to LR simulations (\citealt{2018RAGONE}). BCGs masses are obtained summing over all stellar particles bound to the main subhalo of a group/cluster by Subfind (BCG+ICL) and within a 2D aperture of 50 pkpc.}
        \label{fig:bcgm}
    \end{figure}
    
    In this section we study the properties of our BCGs at $z=0$. In particular we study their mass and their SFR.
    
    \subsubsection{$M_{\star, \bf BCG}-M_{500}$ correlation}\label{sec:mstar-Mbcg}
    
    In Fig.~\ref{fig:bcgm} we show the correlation between the stellar mass of the BCG and $M_{500}$. In this plot, $M_{\star, \rm BCG}$ is defined as the total stellar mass associated to the halo (accordingly to Subfind) and within a $2D$ aperture of $50$ pkpc. We checked that using different line of sight directions brings no more than 40\% difference in the estimated stellar mass, with a median difference of 2\% for our sample. We note that the total mass represents the BCG stellar mass plus the ICL present along the line of sight. In the plot, we also highlights the properties of the BCG that lost its central BH at $z\sim 4$ (see Sect.~\ref{AGN_positioning}).
    
    From the figure we see that our simulations tend to have too massive BCGs with respect to observational data (a factor of $2$ at $M_{500} = 3\times 10^{14}\ M_{\odot}$), in line with the findings of other groups (e.g., \citealt{2017BAHE}, \citealt{2018PILLEPICH}, \citealt{2019HENDEN}). Moreover, current simulations have more massive BCGs than the LR simulations from \cite{2018RAGONE} (orange line in the plot), being a factor of $2.3$ more massive at $M_{500} = 3\times 10^{14}\ M_{\odot}$. A possible explanation could be the different resolution. However, in order to investigate this hypothesis we run a simulation at a $10$ times lower mass resolution and we obtained only a $\sim 12 \%$ lower BCG mass. Hence the BCG mass is very stable against the mass resolution of the simulation (see also \citealt{2018RAGONE}). 
    %This is in line with the analysis of \cite{2018RAGONE}, who ran few test regions at higher resolution than the standard one, pointing out that their results were stable against resolution with no particular trends. 
    To check if the different results with respect to LR simulations (i.e., \citealt{2018RAGONE}) are due to the different implementations of the AGN feedback adopted, we run two more simulations: one at the current mass resolution and the other with a 10 times lower resolution, both implementing the same AGN prescription of \cite{2018RAGONE}. In this set up, gas particles need to be colder than a fixed temperature threshold to be considered multiphase and the energy released by the AGN feedback is used to evaporate the cold gas. Even in this case, we do not find any trend with resolution, the results being in agreement within $5 \%$. Moreover, the BCGs masses obtained in these last two runs are in agreement within $30 \%$ with \cite{2018RAGONE} results (orange line in Fig.~\ref{fig:bcgm}), pointing out that the BCG mass is most sensitive to the prescription adopted for AGN feedback. 
    
    %\GL{frase seguente gia scritta sopra} Hence, the BCG mass is stable against resolution but sensitive to the prescription adopted for AGN feedback. \CR{in order to ascribe the change in BCG mass to the AGN model, shouldn't be compared the BCG mass obtained in the 10x with evaporation+Tthreshod and without?}

    Interestingly, with the \cite{2018RAGONE} set up, together with a lower BCG mass, we also get a lower normalisation for the GSMF, in line with the results of \cite{2019BASSINI}. Thus, we conclude that the different BCG masses are not due to the increased resolution but to the different prescription for the AGN feedback and that, with the current implementation of feedback, it is difficult to simultaneously reproduce both the observed $M_{\rm BCG}-M_{500}$ relation and the GSMF.

    \begin{figure}
        \centering
        \includegraphics[width=\linewidth]{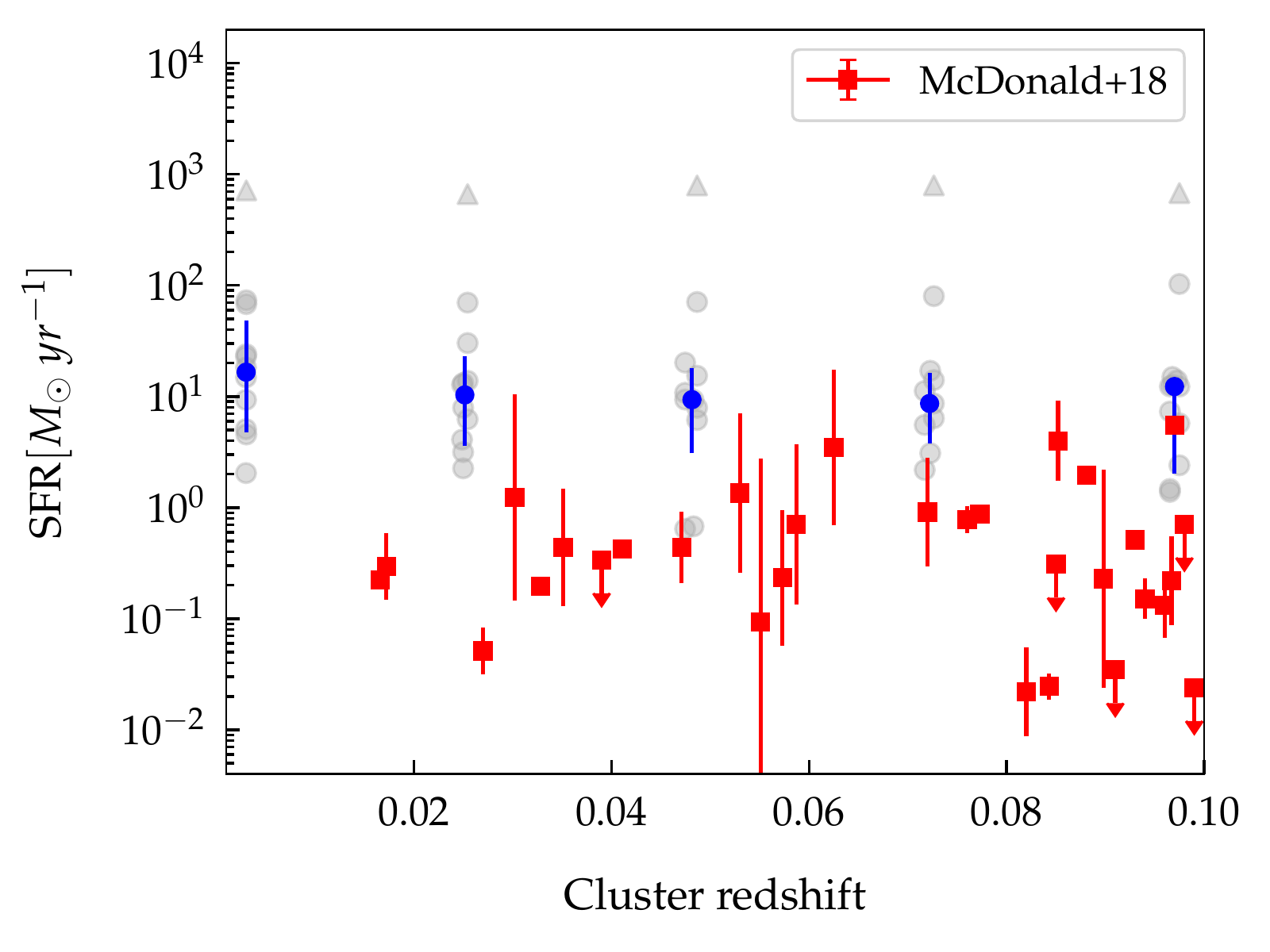}
        \caption{BCG star formation rate in observations and simulations. Grey circles are BCGs of our simulations from different snapshots, while the grey triangle is used for the BCG that lost its central BH at $z \sim 4$ (see Sect.~\ref{AGN_positioning}). BCGs from the same snapshot are shifted only for visualisation purposes. The median values are shown as blue circles and the vertical bars indicate the range between the $16^{\rm th}$ and $84^{\rm th}$ percentiles. A 2D aperture of $30\ \rm pkpc$ is used. Red squares are BCGs from the sample of \cite{2018MCDONALD} (see text for more details).}
        \label{fig:bcg_sfr}
    \end{figure}

    \begin{figure}
        \centering
        \includegraphics[width=\linewidth]{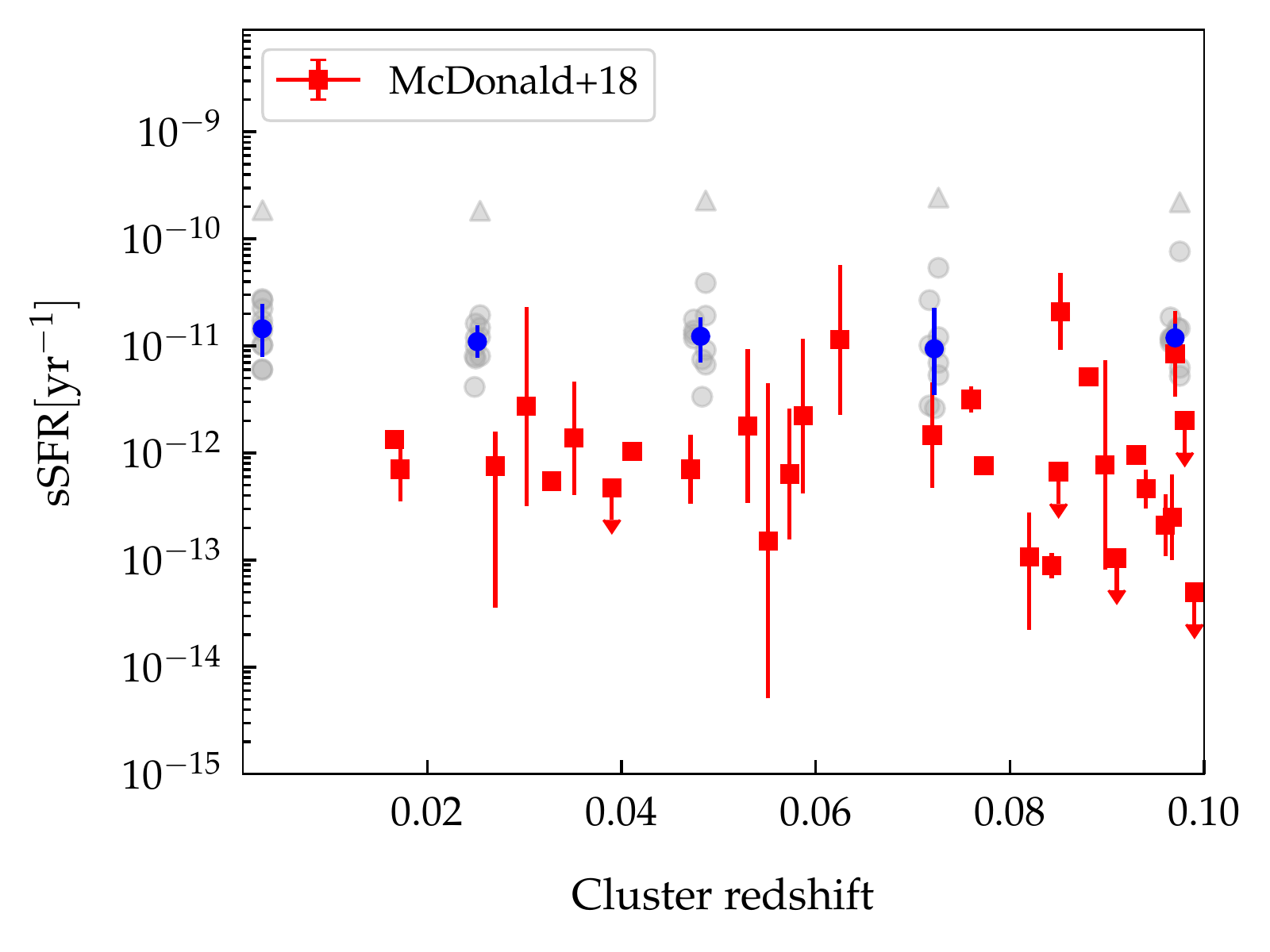}
        \caption{sSFR of BCGs in observations and simulations. Grey circles are BCGs of our simulations from different snapshots (blue circles are median values with $16^{\rm th}$ and $84^{\rm th}$ percentiles), while the grey triangle is used for the BCG that lost its central BH at $z \sim 4$ (see Sect.~\ref{AGN_positioning}). BCGs from the same snapshot are shifted only for visualisation purposes. A 2D aperture of $30\ \rm pkpc$ is used. Red squares are BCGs from the sample of \cite{2018MCDONALD} (see text for more details).}
        \label{fig:bcg_ssfr}
    \end{figure}
    
    \subsubsection{Star formation rate of BGCs}
    
    %\GL{notare, citare e discutere che la SF delle BCG, e non solo a z=0, era stata confrontata con dati pure in Ragone+18. Li avevmo plottato dei dati di hoffer+ 2012 che qui sono ignorati, e darebbero SF residue ben piu' alte. Come mai qui sono ignorati? Sono considerati inaffidabili o che?}
    
    In Fig.~\ref{fig:bcg_sfr} we show the SFR of our simulated BCGs in comparison with observational data. Observations are taken from \cite{2018MCDONALD}, and constitute a subsample of \cite{2014FRASER} BCGs. The original selection has been made considering all galaxy clusters in a volume limited sample, $z<0.1$, with a measured X-ray luminosity in the {\it ROSAT} $0.1-2.4$ keV band $L_X > 10^{44}\ \text{erg s}^{-1}$. This luminosity cut ensures a completeness $> 80\%$ for the cluster sample (see \citealt{2014FRASER}). Moreover, selecting the sample on cluster properties rather than BCG properties enables to correctly account for low values of SFRs and non-detections. However, it has been noted that the SFRs published in \cite{2014FRASER} lack important k-corrections which lead to biased results (see \citealt{2016GREEN}). \cite{2018MCDONALD} recomputed the SFRs using $12\ \mu m$ flux following the procedure of \cite{2016GREEN} for all the clusters with $L_X > 3.3\times 10^{44}\ \text{erg s}^{-1}$. The final sample comprises 33 objects and is complete above the cut in X-ray luminosity. In grey we show the results of our simulations. We emphasise that the BCGs are taken from the same simulated regions at different redshifts, and therefore are not independent. In blue we plot the median value with $16^{\rm th}$ and $84^{\rm th}$ percentiles. To mimic the selection of \cite{2018MCDONALD}, we considered only the $11$ clusters with $M_{500}>2.8\times 10^{14}\ M_{\odot}$ ($M_{200}\gtrsim 4\times 10^{14}\ M_{\odot}$), which corresponds to $L_X > 3.3\times 10^{44}\ \text{erg s}^{-1}$ following the correlation between  $L_X$ and $M_{500}$ showed by \cite{2018TRUONG}. The simulations used in the work of \cite{2018TRUONG} are not the same that we are using for this work, but we checked that using a relation based on our clusters leads to the same final sample. The SFR in simulations is the instantaneous SFR predicted by the effective model for multiphase particles computed considering all the particles bound to the group by Subfind and within a 2D aperture of $30$ pkpc. We employed this aperture to directly compare with other numerical simulations, after checking that the aperture choice does not affect our conclusions. Our simulations present a high residual SFR at these low redshifts ($\sim 10\ \rm M_{\odot}\ \rm yr^{-1}$  against the average observed SFR of $\sim 0.3\ \rm M_{\odot}\ \rm yr^{-1}$). Considering a 3D aperture of $30$ pkpc brings to the same conclusions, as the bulk of SFR is located near the centre of the cluster and the median difference between a 3D and a 2D aperture is $25 \%$. Similar results are also found by other groups. \cite{2019HENDEN} made a similar analysis considering all the clusters with $M_{200}>10^{14} M_{\odot}$ and computed the instantaneous SFR within a spherical aperture of $30\ \rm kpc$ at $z=0.2$. Their results show that their BCGs form stars at a rate similar to ours when they do not have a null SFR (see their Fig.~8). 
    
    Even though the disagreement between simulations and observations is quite large (a factor of $\sim 30$ at $z\sim 0$ according to our results), it is important to keep in mind that measuring the SFR of galaxies is always a non-trivial task, especially in the case of BCGs due to their low SFR values and to the crowded environment. In the particular case of the sample used in our comparison, the SFR is obtained from the $12\ \mu m$ luminosity through the relation derived by \cite{2014CLUVER}. However, this relation is calibrated on star-forming galaxies and flattens at SFR $< 5\ \rm M_{\odot}\ yr^{-1}$. For this reason the values of SFR inferred from observations of BCGs are likely to be underestimated, thus possibly alleviating the tension outlined in Fig.~\ref{fig:bcg_sfr}. The same caution has to be applied to the conclusions drawn in the next paragraph.

    In Fig.~\ref{fig:bcg_ssfr} we plot the sSFR for our simulations and \cite{2018MCDONALD} observations. As we did for the SFR, we checked that the choice of the aperture does not affect our results. Indeed, the results obtained using an aperture of $30$ and $50$ pkpc are in agreement within $30 \%$ at $z=0$. Also in this case, numerical simulations appear to be an order of magnitude above observations. Similar results are also found by other groups. \cite{2019DAVIES} showed that Eagle simulation presents a sSFR $\sim 10^{-11}\ \rm yr^{-1}$ at $M_{200}\sim 10^{14} M_{\odot}$ and $z=0$, and that IllustrisTNG BCGs have a sSFR of $\sim 2.5\times 10^{-12}$ at $M_{200}\sim 10^{14} M_{\odot}$, which is a factor of 3 higher than the median value of \cite{2018MCDONALD} BCGs. Moreover, considering that the BCG sSFR are an increasing function of mass in IllustrisTNG (see Fig.~12 of \citealt{2019DAVIES}) and that the sample of \cite{2018MCDONALD} is of massive clusters ($L_X>3.3\times 10^{44}\ \rm erg\ s^{-1}$), the factor of $\sim 3$ is probably a lower limit.

    %Moreover, IllustrisTNG BCGs sSFR are an increasing function of mass for $M_{200} > 3\times 10^{12}$ (see Fig.~12 of \citealt{2019DAVIES}). Given that \cite{2018MCDONALD} sample is based of clusters with $L_X>3.3\times 10^{44}\ \rm erg\ s^{-1}$, the factor of $\sim 3$ is probably a lower limit.
    
    For our simulations, a possible solution to this mismatch could be a more effective AGN feedback. However, \cite{2018RAGONE} found in their work a very similar result ($\text{sSFR}\sim 1.5\times10^{-11}\ \rm yr^{-1}$ at $z=0$) with an implementation of the AGN feedback more effective in quenching the star formation (see also results in Sect.~\ref{sec:gsmf} and Sect.~\ref{sec:mstar-Mbcg}). Moreover, with the current scheme a more effective feedback could (at least in principle) reduce the gap in the SFR between simulated and observed BCGs, at the price of worsening the difference in the GSMF (see Sect.~\ref{sec:gsmf}). Therefore, a better solution would be to modify the prescription of the AGN feedback, in order to have more efficient quenching only for massive galaxies.
    
    %with the current implementation of the feedback, increasing the efficiency would affect other observable with consequent mismatch with observations (such as, for example, the Magorrian relation showed in Fig.~\ref{fig:magorrian}). Several studies (\cite{2016GASPARI} for a brief review)showed that in massive galaxies the physical process leading the accretion into SMBHs is the chaotic cold accretion (CCA). while our current implementation of the feedback mimic CCA in power, it does not capture the rapid frequency of chaotic events. CCA leads to the rapid funneling of cold clouds also at low redshift, as the hot halo develops nonlinear thermal instability and a consequent rain of clouds toward the inner SMBH. Such frequent accretion, key to quench cooling flows, showld solve the differences highlighted in Fig.~\ref{fig:bcg_sfr} and Fig.~\ref{fig:bcg_ssfr}. \GL{Mi sembra molto speculativo, Almeno bisognerre spiegare con piu' dettaglio perche' ci si aspetta questo. A me non risulta ovvio. O almeno non è ovvio perche' migliorebbe questo aspetto senza danneggiarne altri, analogamente all'aumento della efficienza}
%--------------------------------------------------------------------

\section{Protoclusters at $z\sim 2$}\label{sec:proto_z2}

    %From our simulations, we define `protocluster' as the core of the Lagrangian region that will eventually collapse into a cluster at $z=0$, with a typical size of $\sim 1$ pMpc, instead of the entire Lagrangain region, which is up to few tens of Mpc at $z\sim2$ (see \citealt{2015MULDREW}, \citealt{2018BIFFI}).
    
    %In particular, we compare with the observations of \cite{2014CLEMENTS}, who studied the integrated SFR of 4 protocluster regions, and \cite{2016WANG} and \cite{2019GOMEZ}, who studied both the integrated SFR and the contribution from single resolved sources.
    In this section we compare our simulations to observational results of protocluster regions of different sizes and identified at $z\sim 2$.  Since we are interested in comparing values of SFR, we include only protocluster regions with coverage in the FIR and sub-millimetric bands, as the total star formation budget is mainly contributed by its obscured component. 
    In particular, we compare with the observations of \cite{2014CLEMENTS}, \cite{2014DANNERBAUER}, \cite{2016WANG}, \cite{2016KATO}, \cite{2018COOGAN}, and \cite{2019GOMEZ}.
    \subsection{Protocluster SFR within $\sim 1 \ \rm pMpc$} 
    
    \begin{figure}
        \centering
        \includegraphics[width=\linewidth]{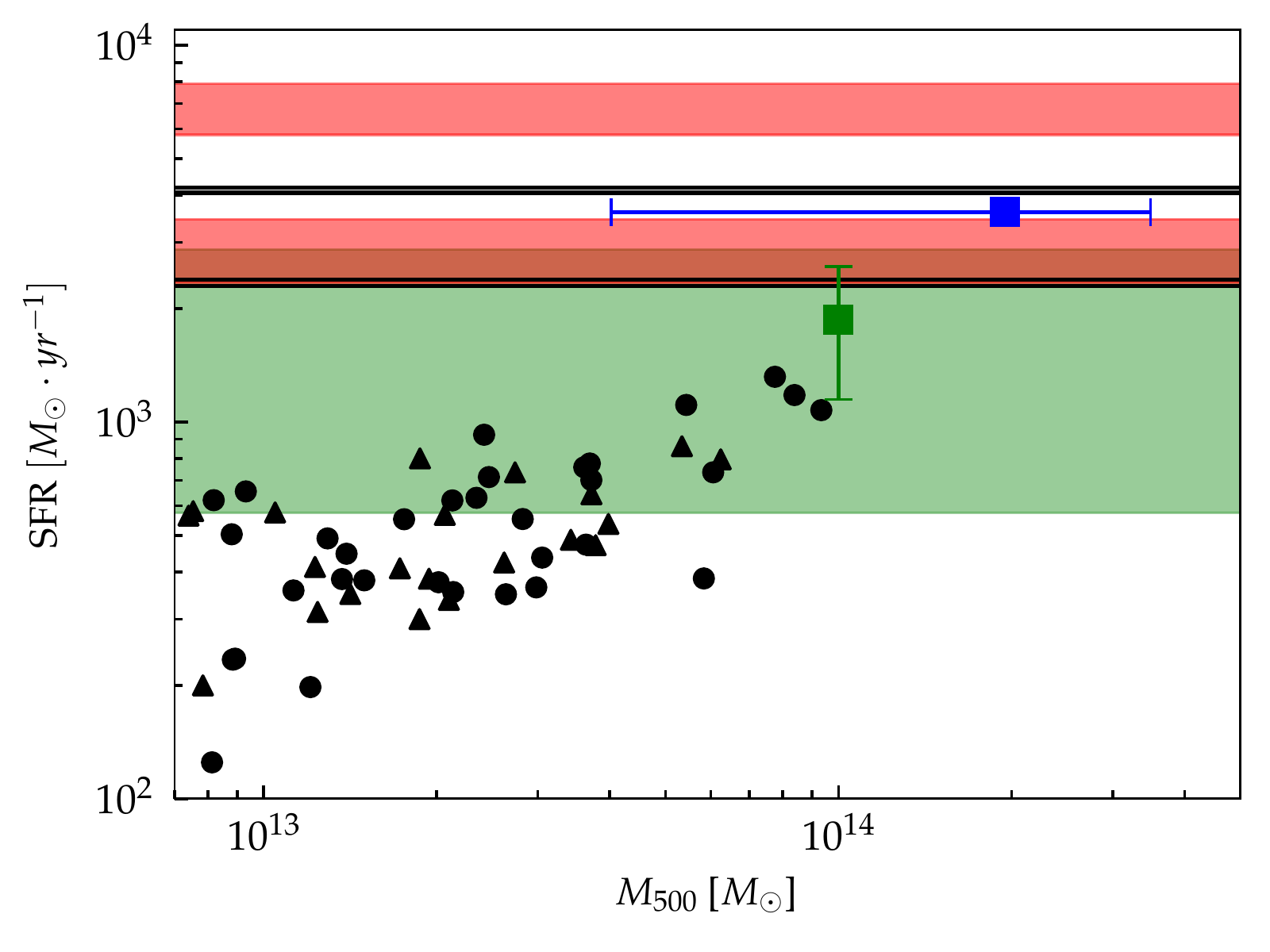}
        \caption{SFR of protocluster regions at $z\sim 2$ in observations and simulations within an aperture of $\sim$ 2 pMpc. Red bands refer to two clumps from \cite{2014CLEMENTS}, black solid lines refer to four fields from \cite{2010STEVENS} and analysed by \cite{2014CLEMENTS}. Blue square refer to the Spiderweb structure (\citealt{2014DANNERBAUER}). Green square and green band refer to the two protoclusters analysed by \cite{2016KATO} (HS1700 and 2QZCluster respectively). Black circles and triangles refer to numerical simulations, where the SFR is plotted against protocluster mass (see text). We used black circles for groups which end up in the central cluster of the region at $z=0$, and black triangles otherwise.}
        \label{fig:m500_sfr}
    \end{figure}

    In Fig.~\ref{fig:m500_sfr} we compare the observed star formation obtained by \cite{2014CLEMENTS}, \cite{2014DANNERBAUER}, and \cite{2016KATO} with the same quantity computed in simulations. With the red bands we show the protocluster regions at $1\lesssim z\lesssim 3$ identified as clumps in Planck $857$ GHz band by \cite{2014CLEMENTS}. This frequency is suitable for identifying DSFGs, which trace star-bursting phases of protoclusters (e.g., \citealt{2004GRANATO}). They retrieve the far-infrared luminosity, $L_{\rm FIR}$, by fitting the spectral energy distribution (SED) with a modified black body formula, and then they compute the SFR using the relation given by \cite{2003BELL}, which assumes a Salpeter IMF (\citealt{1955SALPETER}). In order to compare with our simulations, which instead adopt a Chabrier IMF (\citealt{2003CHABRIER}), we divide their values of SFR by $1.74$.  We also include in our comparison the analysis that \cite{2014CLEMENTS} performed on the fields previously introduced by \cite{2010STEVENS}, who analysed the density flux of submillimeter galaxies (SMGs), obtained at $850\ \mu \rm m$ in 5 fields centred on QSOs in the redshift range $1.7<z<2.8$. The SFRs for these fields are computed following the procedure already described, with the difference that the FIR luminosity is computed from the $F_{850}$ flux using an Arp 220 spectral template. Also in this case we corrected for the choice of the IMF. Among the 4 clumps and 5 fields analysed in \cite{2014CLEMENTS} we include in our comparison the 6 residing at $1.74 < z < 2.27$ (the values of SFR for the four fields are 2240, 2330, 3966, and 4095 $\rm M_{\odot}\ \rm yr^{-1}$, and thus they overlap in Fig.~\ref{fig:m500_sfr}). The physical volume used to compute the SFRs in \cite{2014CLEMENTS} clumps is $4.2\ \rm Mpc^3$, equal to a sphere of $1\ \rm Mpc$ radius, while the fields are characterised by a volume of $1.4\ \rm Mpc^3$, equivalent to a sphere of $0.7\ \rm Mpc$ radius. 
    
    \cite{2014DANNERBAUER} studied the FIR properties of the protocluster associated to the radio-galaxy HzRG MRC1138-262 at $z = 2.16$ (also known as Spiderweb galaxy, \citealt{2000PENTERICCI,2006Miley}). This structure is characterised by an overdensity of Lyman alpha emitters (LAEs), and has been studied in different sub-millimeter wavelengths (see \citealt{2014DANNERBAUER} and references there in). Observations in the FIR (100, 160, 250, 350, 500, and 850 $\mu m$) were used to fit the SED of detected sources assuming a grey body formula, which was used to compute the total FIR luminosity and the correlated SFR through the relation given by \cite{1998KENNICUTT}. The resulting SFR computed within a sphere of $1\ \rm pMpc$ radius, corrected to a Chabrier IMF, is $\sim 3600\ \rm M_{\odot}\ yr^{-1}$ and is showed as a blue square in Fig.~\ref{fig:m500_sfr}. Numerical simulations suggest that this protocluster is the progenitor of a massive $z=0$ cluster (\citealt{2009SARO}), with a predicted mass of $M_{200}\sim 10^{15}\ \rm M_{\odot}\ yr^{-1}$. Therefore, this structure is a candidate progenitor of the massive simulated clusters used in this study. Finally, we also show the observations in the FIR of other two already known protoclusters: 2QZCluster ($z=2.23$, \citealt{2011MATSUDA}) and HS1700 ($z=2.3$, \citealt{2005STEIDEL}). \cite{2016KATO} used SPIRE bands (250, 350, and 500 $\mu m$) to obtain a colour-selected sample of DSFGs possibly associated to these two protoclusters. In their work, they found overdensities of DSFGs in both protoclusters regions, even though the redshift of these sources is not confirmed yet. Assuming a grey-body spectrum, they computed $L_{\rm IR}$ in the $\sim 1\ \rm pMpc$ region containing the highest number of DSFGs obtaining values very similar to the one reported by \cite{2014CLEMENTS}. We show \cite{2016KATO} results in Fig.~\ref{fig:m500_sfr} as a green square (HS1700) and green band (2QZCluster). In both cases the upper limit on the value of SFR is obtained assuming that all the detected sources are within the protocluster, while the lower limit is obtained subtracting field average values (see \citealt{2016KATO} for further details).

   Regarding the data from simulations, we considered at $z=2$ the five most massive groups identified by Subfind in each of the analysed regions. The mass $M_{500}$ of the group is given by Subfind, while the SFR is the sum of the instantaneous SFR of all gas particles within a sphere of $1\ \rm Mpc$ radius from the centre of the group.  This aperture matches the volume adopted for \cite{2014CLEMENTS} clumps, and is slightly larger than the volume of \cite{2010STEVENS} fields. In addition, We also adopt another possible definition of the SFR: the SFR averaged over $\sim 100\ \rm Myr$. Although this may be the optimal choice when comparing to DSFGs, it does non quantitatively affect our results as our most star forming protocluster region is characterised by a $2\%$ difference in the SFR when the two methods are used. For this reason we only use the instantaneous SFR throughout the paper. In Fig.~\ref{fig:m500_sfr} we also differentiate between the progenitors of the most massive cluster at the centre of each region by $z=0$ and the groups that will form other objects. The former are plotted as black circles, the latter as black triangles. From the plot it is clear that the simulated protoclusters do not reproduce the high SFR observed, as the difference between the highest value of SFR rate reached within our set of simulated protoclusters ($\sim 1300\ \rm M_{\odot}\ yr^{-1}$) is a factor of $\sim 5$ lower than the SFR measured in one of the \cite{2014CLEMENTS} clumps ($\sim 7000\ \rm M_{\odot}\ yr^{-1}$), and a factor of $\sim 3$ lower than the SFR measured within the protocluster associated to the Spiderweb galaxy ($\sim 3600\ \rm M_{\odot}\ yr^{-1}$). This result is in agreement with the conclusions of \cite{2015GRANATO}, who used a set of simulations with the same initial conditions used here but at 10 times lower mass resolution and a previous version of our code (the main difference being the prescription for AGN feedback and BH repositioning, see \citealt{2013RAGONE}) together with dust reprocessing and radiative transfer post-processing performed with GRASIL-3D (\citealt{2014DOMINGUEZ}) to directly compare FIR fluxes with \cite{2014CLEMENTS}. \cite{2015GRANATO} concluded that simulations fail to reproduce the observed fluxes at $z=2$ by a factor $\gtrsim 3-4$. Given the results showed in Fig.~\ref{fig:m500_sfr}, we conclude that the results of \cite{2015GRANATO} hold at higher resolution and are not dependent on the particular prescription adopted for AGN feedback.
    %are not affected by numerical resolution.
    
    \begin{figure}
        \centering
        \includegraphics[width=\linewidth]{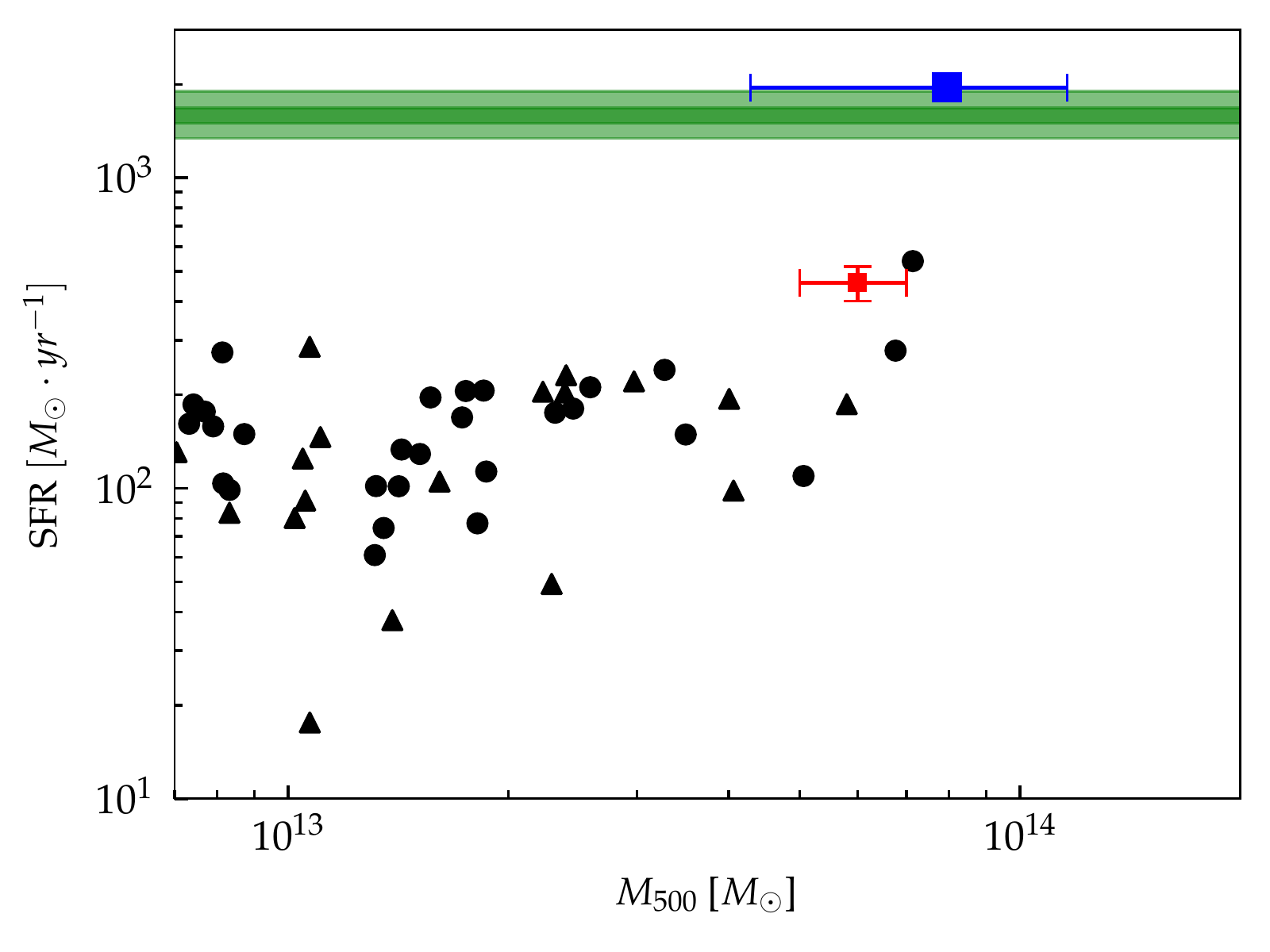}
        \caption{SFR of protocluster regions at $2<z<2.6$ in observations and simulations within an aperture of $\sim$ 100 pkpc. Green bands refer to two protoclusters from \cite{2019GOMEZ}, blue square refers to \cite{2016WANG}, and red square refers to \cite{2018COOGAN}. Black circles and triangles refer to numerical simulations, where the SFR is plotted against protocluster core mass. We used black circles for groups which end up in the central cluster of the region at $z=0$, and black triangles otherwise.}
        \label{fig:m500_sfr_z2.4}
    \end{figure}
    
    \begin{figure}
        \centering
        \includegraphics[width=\linewidth]{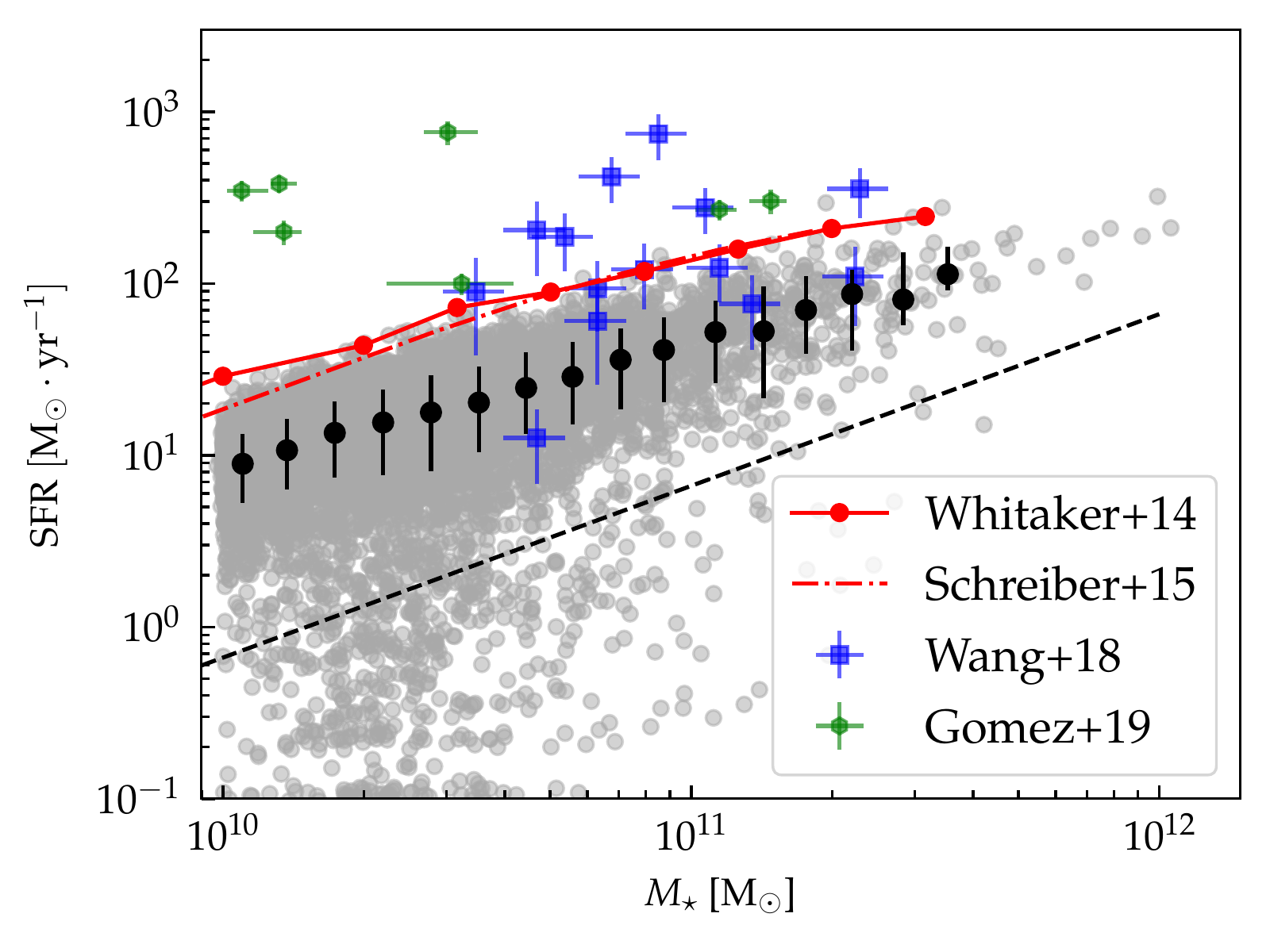}
        \caption{SFR as a function of galaxy stellar mass at $z\sim 2.3$. Red solid and dashed lines are observational data from \cite{2014WHITAKER} and \cite{2015SCHREIBER} respectively. Green hexagons and blue squares are galaxies from the protoclusters of \cite{2019GOMEZ} and the cluster of \cite{2018WANG} respectively. Grey points are galaxies in our simulations. Black dashed line fix the distinction between active and passive galaxies (\citealt{2016PACIFICI}). Black points represent median values of star forming galaxies with 16th and 84th percentiles. Both SFRs and stellar masses are computed considering a 3D aperture of $30\ \rm pkpc$.}
        \label{fig:mstar_sfr}
    \end{figure}

    \begin{figure}
        \centering
        \includegraphics[width=\linewidth]{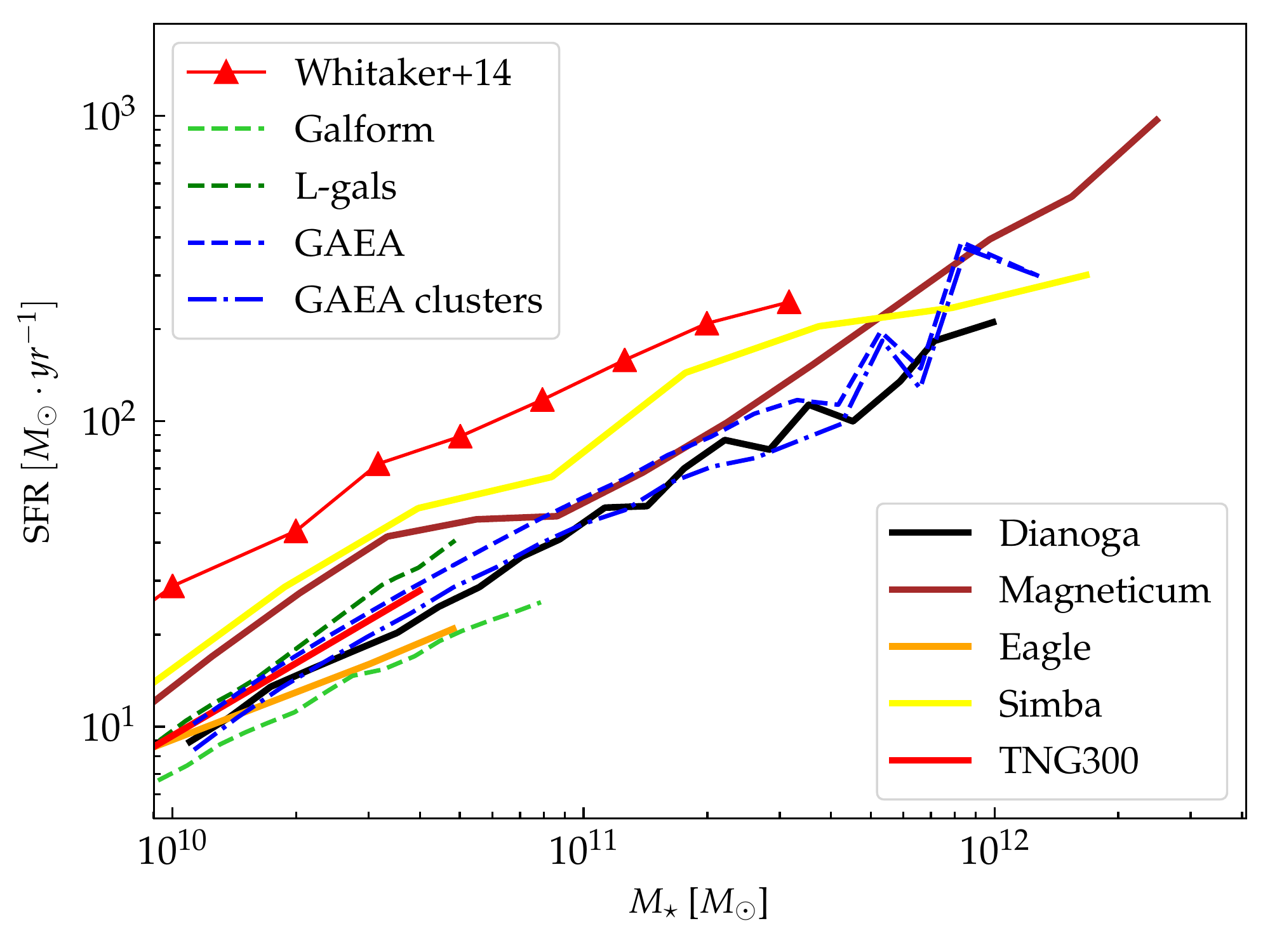}
        \caption{Main sequence of star forming galaxies at $z\sim2$. Red triangles are observational data from \cite{2014WHITAKER}. Black line are median values for our simulations. Coloured solid and dashed lines are data from other cosmological simulations and semi-analytical models respectively. In particular: Eagle (orange solid line, \citealt{201GUO}), TNG300 (red solid line, \citealt{2019DONNARI}), Simba (yellow solid line, \citealt{2019DAVE}), Galform (green dashed line, \citealt{201GUO}), L-galaxies (dark green dashed line, \citealt{201GUO}), and GAEA (blue dashed line, \cite{2014HIRSCHMANN}). For the GAEA model we also show the results obtained considering only galaxies that at $z=0$ are within galaxy clusters with mass $>10^{14.25} \rm M_{\odot}$ (see text for more details).}
        \label{fig:mstar_sfr_sim}
    \end{figure}
    
    \subsection{Protocluster SFR within $\sim 100 \ \rm pkpc$}
     
    In Fig.~\ref{fig:m500_sfr_z2.4} we compare simulations with observations by \cite{2016WANG} (blue square), \cite{2018COOGAN} (red square), and \cite{2019GOMEZ} (green bands). 
    %In the case of \cite{2016WANG} and \cite{2019GOMEZ} observations were complemented with ALMA data with the consequent possibility to resolve single galaxies in the protocluster cores. 
    In particular, \cite{2016WANG} recently discovered a cluster (CL J1001+0220) at $z=2.506$. This structure, detected as an overdensity of Distant Red Galaxies (DRGs), appears as a massive, virialized halo. Through an analysis of the velocity dispersion, stellar mass content, and also the detected X-ray emission \cite{2016WANG} estimated the cluster mass to be $10^{13.9 \pm 0.2}M_{\odot}$. However, differently from local clusters, CL J1001+0220 is characterised by a high fraction of massive ($M_{\star}>10^{11} M_{\odot}$) star forming galaxies. The SFR in the $80\ \rm pkpc$ core region, computed from FIR luminosity and corrected to a Chabrier IMF, is estimated to be $\sim 2000\ M_{\odot}\ \rm yr^{-1}$. Moreover, the fraction of starbursting galaxies is $\sim 25\%$, much higher than the value in the field which is about 3\% accordingly to \cite{2015SCHREIBER}. \cite{2019GOMEZ} spectroscopically confirmed two protoclusters through CO emission lines (a third one has been analysed in the same work, but it was associated with the well known CL J1001+0220 cluster in the COSMOS field, see \citealt{2016WANG}, \citealt{2018WANG}). These objects were previously recognised as separate sources by \cite{2015BUSSMANN} who observed with ALMA 29 bright DSFGs taken from the Hermes Survey (\citealt{2012OLIVER}). The two new protoclusters are composed of $4$ and $5$ gas rich DSFGs over a region of $125\ \rm pkpc$ and $64\ \rm pkpc$ at $z=2.171$ and $z=2.602$ respectively. $L_{\rm IR}$, used to compute the SFRs of single sources following \cite{1998KENNICUTT}, are computed integrating the SED fitted from the IR available fluxes measurements at 24, 250, 350, 500 $\mu m$, and 3 mm.
    
    Finally, we include in the comparison also the observations by \cite{2018COOGAN} of the protocluster Cl J1449+0856, identified by \cite{2011GOBAT} as an overdensity of IRAC colour-selected galaxies. The protocluster has also been detected from the X-ray emission, from which it has been estimated a mass in the range $[4-6]\times 10^{13}\ \rm M_{\odot}$ (\citealt{2016VALENTINO}). \cite{2018COOGAN} employed ALMA observations of $870\ \mu m$ continuum and CO(4-3) emission line to compute the SFR within the $\sim 0.08\ \rm pMpc^2$ cluster central region (see \citealt{2018COOGAN} or \citealt{2018STRAZZULLO} for more details on the computation of the SFR). The value of SFR reported in fig.~\ref{fig:m500_sfr_z2.4} is the sum of the SFR obtained for obscured ($\sim 400\ \rm M_{\odot}\ yr^{-1}$) and unobscured ($\sim 60\ \rm M_{\odot}\ yr^{-1}$) star formation. As a final remark, we note that the values of SFR computed through the $870\ \mu m$ continuum rely on the assumed SED template. The values reported in Fig.~\ref{fig:m500_sfr_z2.4} are obtained considering a template for main sequence galaxies. Considering a template typical of starburst galaxies, the value of SFR would be twice as high (e.g., \citealt{2018STRAZZULLO}).
    
    In numerical simulations we computed the instantaneous SFR considering a 2D aperture of $90\ \rm pkpc$ (around the mean value of the four observed protoclusters that we use as comparison) and integrating $1\ \rm pMpc$ along the line of sight. The choice of the projected distance does not affect our results as most of the stars are produced at the centre of the protoclusters. Indeed, we verified that the results are quantitatively the same integrating up to $3\ \rm pMpc$ along the line of sight. Similarly to previous results, the most star forming region within our set of simulations (SFR $\sim 500\ M_{\odot}\ \rm yr^{-1}$) underpredicts the highest observed SFR (SFR $\sim 2000\ M_{\odot}\ \rm yr^{-1}$) by a factor of $\sim 4$. 
    
    %Finally, from Fig.~\ref{fig:m500_sfr} we can see that simulations predict a positive correlation between the mass of a protocluster and the total SFR. Indeed, the Pearson coefficient for the correlation ${\rm Log}\ M_{500}- {\rm Log}\ SFR$ is $r=0.67$. However, in Fig.~\ref{fig:m500_sfr_z2.4}, where the SFR is computed within a smaller aperture ($[0.3, 0.5]\times R_{500}$) the correlation is weaker with a correlation coefficient $r=0.34$. Hence, simulations predict a correlation between cluster mass and integrated SFR, when the SFR is computed within an aperture $\gtrsim R_{500}$. (NOTE FOR MYSELF: WRITE AND CONTEXTUALiZE BETTER THIS LAST ARGUMENT)

    \subsection{Main sequence of star forming galaxies}
    
    To explore a possible origin for the difference between SFRs in simulations and observations, in Fig.~\ref{fig:mstar_sfr} we show the observed and simulated correlation between stellar mass and SFR in galaxies. Red solid line represents the main sequence (MS) of star forming galaxies as derived by \cite{2014WHITAKER}, obtained considering star forming galaxies selected in UVJ colours in the redshift range $2<z<2.5$. The red dashed line is the main sequence at $z\sim 2.3$ from \cite{2015SCHREIBER}, who used a similar approach but considered only photometry at rest-frame wavelengths larger than $30\ \rm \mu m$ to avoid pollution from AGNs. To compare with our simulations and \cite{2014WHITAKER} we corrected \cite{2015SCHREIBER} main sequence to a Chabrier IMF. Green hexagons and blue squares are single galaxies from the protoclusters of \cite{2019GOMEZ} and the cluster of \cite{2018WANG} respectively. Grey points represent all the simulated galaxies in our regions at $z\sim 2.3$. The dashed black line is a redshift dependent threshold in sSFR which distinguishes quiescent from star forming galaxies (\citealt{2016PACIFICI}) and mimics the selection in UVJ colours while the black points with errorbars are median values with $16^{\rm th}$ and $84^{\rm th}$ percentiles of simulated star forming galaxies. 
    
    As we can see from the plot, the SFRs of simulated galaxies are below the observed relation by a factor of $\sim 3$. This is a known discrepancy between simulations (and also semi-analytical models) and observations (see, for example, \citealt{2016DAVE} and \citealt{2017XIE}). Indeed, around the peak of the cosmic star formation rate density simulations show a normalisation for MS of star forming galaxies that is typically a factor of $\sim 2-3$ lower than observations. In Fig.~\ref{fig:mstar_sfr_sim} we show the results from different numerical simulations and semi-analytical models. All but our simulations refer to cosmological boxes, so that a possible bias in our results toward a lower main sequence normalisation can be expected when comparing with other simulations. However, observational works do not find differences between the MS computed in different environments (e.g., \citealt{2013KOYAMA}). This result is also in line with the predictions of the GAEA semi-analytical model (\citealt{2016HIRSCHMANN}). 
    %SB. QUesto è un risultato che hai calcolato tu o te lo ha passato Fontanot? Nel secondo caso lo dovresti citare come Fontanot (private communication) e poi ringraziarlo esplicitamente per questo negli acknowldgements. 
    In particular, we computed the MS of star forming galaxies considering only the main progenitors of the galaxies that are found in a cluster with mass $M_{\rm vir}>10^{14.25}\ \rm M_{\odot}$ at $z=0$ (GAEA clusters in the plot), finding no more than a $30\%$ difference with respect to the MS obtained considering all active galaxies in the simulation.
    
    The discrepancy outlined in Fig.~\ref{fig:mstar_sfr_sim} is an interesting feature given the fact that this difference persists also in numerical simulations which reproduce the GSMF at every redshift (e.g., \citealt{2019DAVE}). A systematic factor of $\sim 3$ in the galaxy star formation rate at $z\sim 2$, that naturally arises in case the observed normalisation of the main sequence is matched, could alleviate the discrepancy between SFR in simulated and observed protocluster (see Fig.~\ref{fig:m500_sfr} and Fig.~\ref{fig:m500_sfr_z2.4}). Moreover, looking at individual galaxies in observed protocluster regions in Fig.~\ref{fig:mstar_sfr}, we see that most of them are above the main sequence with also few galaxies classified as starburst. In this respect it is interesting to note that galaxies within the cluster identified by \cite{2016WANG}, detected also in X-ray and hence probably in a more mature evolutionary stage with respect to the structures identified by \cite{2019GOMEZ}, are characterised by higher masses and are scattered around the observed main sequence. On the contrary, galaxies within \cite{2019GOMEZ} structures have lower masses and very high SFRs, all above the main sequence. The level of SFR of these galaxies is not reproduced by simulations, that, besides underpredicting the normalisation of the main sequence, do not exhibit strong starbursts (see Sect.~\ref{sec:discussion_sb}).

    %and in the cluster by \cite{2016WANG}, the galaxies at the center of the structure identified by \cite{2016WANG} are more massive with a fraction below the main sequence. Being the structure observed by \cite{2016WANG} more mature and classified as a cluster, this difference may suggest an evolutionary scenario in which young galaxies evolve at almost constant SFR above of and finally on the main sequence, finally being quenched once they exhaust their gas reservoir. This view is supported by the short gas depletion time for \cite{2016WANG} cluster core, estimated to be $\sim 200\ \rm Myr$. 
    
 %--------------------------------------------------------------------
 \section{Protoclusters at $z\sim 4$}\label{sec:proto_z4}
    
    In this section we compare our simulations to observational results of protocluster regions identified at $z\sim 4$. In particular we compare with the observations by \cite{2018OTEO} and \cite{2018MILLER}. Before digging into our results, it is important to make few considerations. First, the protoclusters studied at $z \sim 2$ in the previous Section come from relatively small surveys, the largest being the one analysed by \cite{2014CLEMENTS}.
    This survey encompasses 90 deg$^2$, that in the redshift range $0.76 < z < 2.3$ corresponds to $\sim 0.6\ h^{-3}\ \rm cGpc^3$ in our cosmology, and thus is smaller than the cosmological box from which the simulated clusters are extracted (see also \citealt{2015GRANATO}). This is not true for the protoclusters studied by \cite{2018OTEO} and \cite{2018MILLER}. The first has been identified within the H-ATLAS fields, corresponding to a total sky area of $\sim 600\ \rm deg^2$ (\citealt{2016IVISON}), while the second comes from a catalog from $\sim 770\ \rm deg^2$ of the South Pole Telescope Sunyaev-Zel'dovich (SPT-SZ) survey (\citealt{2013MOCANU}). The comoving volume corresponding to the H-ATLAS fields in the redshift range $2.7<z<6.4$, corresponding to the redshift spanned by the ultra-red galaxies selected with Herschel by \cite{2016IVISON}, is $\sim 10\ h^{-3}\ \rm cGpc^3$, about ten times larger than the box from which the simulated clusters are extracted. Therefore, \cite{2018OTEO} and \cite{2018MILLER} structures could be sufficiently rare not to be sampled by our simulations. Moreover, the protoclusters analysed at $z \sim 2$ include few bona fide $z=0$ massive clusters (\citealt{2014DANNERBAUER}, \citealt{2016WANG}, \citealt{2018COOGAN}). However, this may not be the case for the ones observed by \cite{2018OTEO} and \cite{2018MILLER}, as it is not guaranteed that a halo of $M_{\rm halo} \sim 10^{13}\ M_{\odot}$ at redshift $\sim 4$ will eventually evolve to a Coma-like structure at $z=0$. Indeed, numerical simulations suggest that the value of the mass of the most massive halo in a protocluster region at $z\sim 4$ is not enough to safely predict the cluster mass by $z=0$. An analysis of the large scale structure, such as the galaxy overdensity over a scale of $\sim 5\ \rm pMpc$, would be needed to place better constraints on the final cluster mass (see \citealt{2013CHIANG}). It is important to keep this in mind when comparing observations and simulations.
    
    %Finally, it is important to remark that we base our comparison on the assumption that SPT2349-56 and the protocluster observed by \cite{2018OTEO} are the progenitors of massive clusters. While we know that our simulated protoclusters will end up in massive clusters by $z=0$,  this may not be the case for the observed ones, as it is not guaranteed that a halo of $M_{\rm halo} \sim 10^{13}\ M_{\odot}$ at redshift $\sim 4$ will eventually evolve to a Coma-like structure at $z=0$. Indeed, numerical simulations suggest that the value of the mass of the most massive halo in a protocluster region at $z\sim 4$ is not enough to safely predict the cluster mass at $z=0$.  An analysis of the large scale structure, such as the galaxy overdensity over a scale of $\sim 5\ \rm pMpc$, would be needed to place better constraints on the final cluster mass (see \citealt{2013CHIANG}).
    
\begin{figure}
        \centering
        \includegraphics[width=\linewidth]{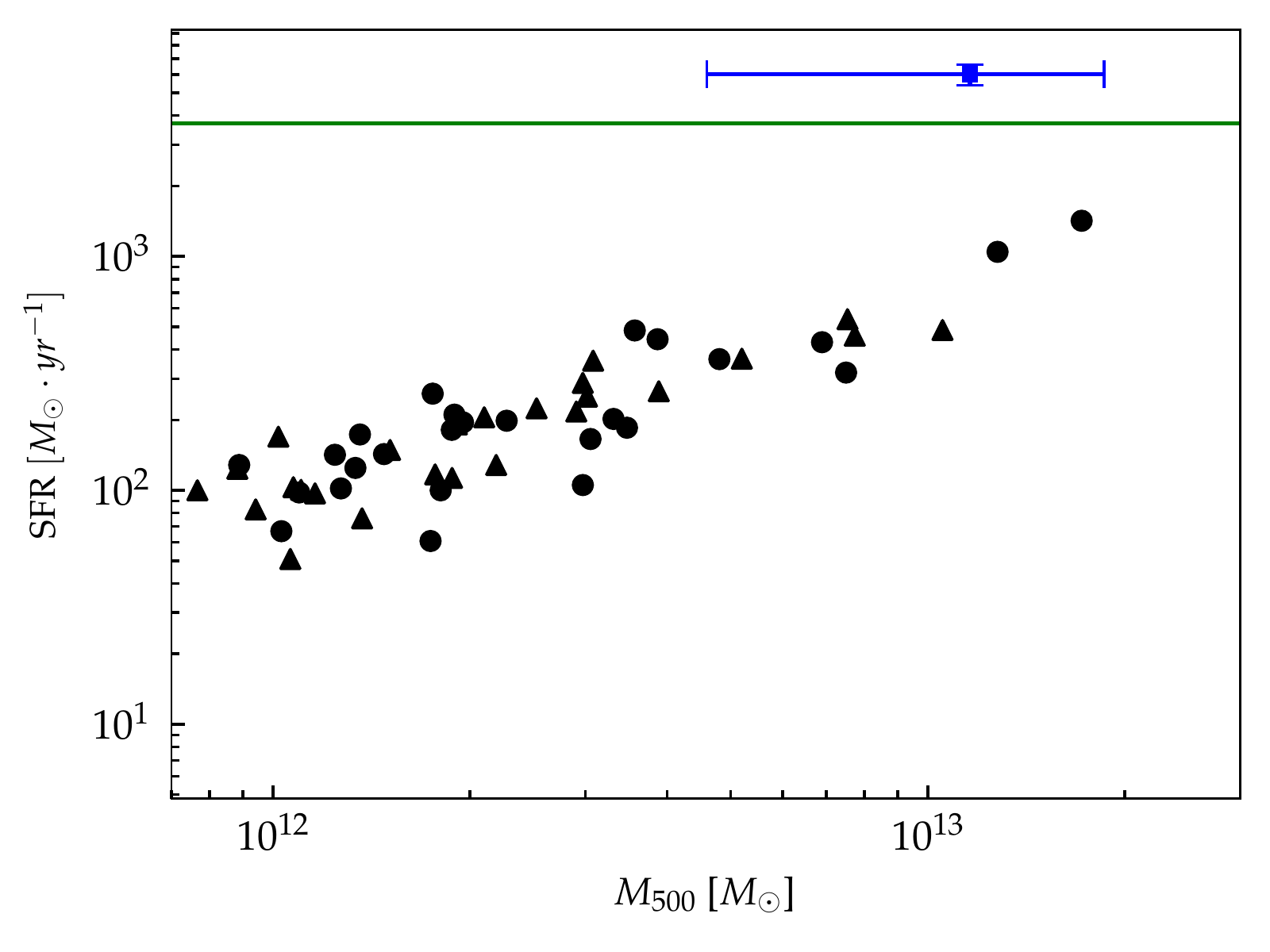}
        \caption{SFR as a function of $M_{500}$ at $z\sim4.3$. Blue square and green line are the observed values of \cite{2018MILLER} and \cite{2018OTEO}
        protoclusters respectively. Black symbols refer to the SFR computed in a cylinder 1 pMpc long and within a circular aperture of $130\ \rm pkpc$ in our simulations. The five most massive groups of each region are shown. We used black circles for groups which end up in the central cluster of the region at $z=0$, and black
        triangles otherwise.}
        \label{fig:protoz4_sfr}
    \end{figure}

    The observations by \cite{2018OTEO} and \cite{2018MILLER} of two highly star forming protocluster cores at $z \sim 4$ and $z \sim 4.3$ are shown in Fig.~\ref{fig:protoz4_sfr} with a green line and blue squares respectively. The protocluster core presented by \cite{2018OTEO} was firstly detected as part of an overdensity of DSFGs in the wide-field LABOCA (a low resolution bolometer camera on the APEX telescope)  map at $870\ \mu \rm m$. Subsequent observations with ALMA at $2\ \rm mm$ and $3\ \rm mm$ revealed that the most luminous source consists of at least $11$ separate sources, of which $10$ were spectroscopically confirmed to be at $z=4.002$ and thus are part of the same structure. $L_{\rm IR}$ for the ALMA resolved sources are computed considering the flux density at $2\ \rm mm$ ($\sim 400\ \mu \rm m$ at rest frame) and assuming an ALESS template for the SED. The resulting SFR (corrected to a Chabrier IMF) in the $260\ \rm pkpc \times 310\ \rm pkpc $ central region is $\sim 3700\ \rm M_{\odot}\ yr^{-1}$. We note that this value is highly uncertain as it depends on the assumed SED template. However, \cite{2018OTEO} showed that within a large range of SED templates, only the one reported by \cite{2013PEARSON} yields a lower SFR than the one obtained with ALESS (by a factor of $0.66$). The protocluster from \cite{2018MILLER}, SPT2349-56, has been first detected by the South Pole Telescope (SPT). Subsequents follow up with LABOCA and ALMA allowed to identify 14 sources in an extremely small area ($\sim 130$ kpc diameter) at $z=4.3$. SFRs are derived from $870\ \mu$m flux density ($S_{870\ \mu m}$) assuming a SFR-to-$S_{870\ \mu m}$ ratio of $150 \pm 50\ \rm M_{\odot}\ \rm yr^{-1} / \rm mJy$, which is typical for SMGs. The gas mass of all 14 galaxies is computed from CO(4-3) line luminosity (converted to CO(1-0) line luminosity through the ratio $r_{4,1}=0.41\pm 0.07$) assuming a $\rm CO/H_2$ conversion factor of $\alpha_{\rm CO} = 0.8\frac{\rm M_{\odot}}{\rm K\ km\ s^{-2}\ pc^2}$ and through the relation:
    \begin{equation}\label{eq:CO}
        M_{\rm gas} = \alpha_{\rm CO} L'_{\rm CO(1-0)}.
    \end{equation}
    When CO(4-3) line is not detected, [CII] line luminosity is converted to CO(4-3) using the average CO(4-3)/[CII] ratio for their detected sample (\citealt{2018MILLER}). 
 
     SFR in simulations is computed considering a 2D aperture of $130\ \rm pkpc$ and integrating $1\ \rm pMpc$ along the line of sight. We also checked that the choice of the length of the cylinder does not quantitatively affect our results. In particular, integrating over the whole box along the line of sight and using different orientations for the cylinder axis induce differences in the measured SFR not higher than $50\%$ at SFR $> 400\ \rm M_{\odot}\ \rm yr^{-1}$. This suggests that at this redshift, the star formation takes place only in the densest simulated regions. As we have done at $z \sim 2$, for each of our regions we selected the $5$ most massive groups at $z=4.3$. Also in this case, the most star forming group within our simulations differ from observations by a factor of $\sim 4$. 
     
     It is, however, important to note that both SPT2349-56 and the protocluster observed by \cite{2018OTEO} represent really rare objects. In fact, none of our simulated protoclusters at $z\sim 4$ has more than $7$ star forming galaxies with a mass higher than $10^{10}\ \rm M_{\odot}$, while \cite{2018OTEO} and \cite{2018MILLER} spectroscopically confirmed 10 and 14 sources respectively. Thus, we conclude that among the main progenitors of our $12$ clusters ($7$ of which very massive), we do not have a structure with the same number of star forming galaxies as these observed protoclusters. Even though this can certainly be due to the limited statistics of the simulated volumes, it can also be  related to the star formation subgrid model, which does not correctly describe galaxy properties at this redshift, or both. Nevertheless, assuming that doubling the number of star forming galaxies within a protocluster to match the observed number within SPT2349-56 would also double the total SFR, the SFR of the 'boosted' simulated protoclusters would still be a factor of $\sim 2$ lower than the observed SFR.
     
     %at this redshift we predict a difference between observations and simulations of a factor of $\sim 2$.

    \begin{figure}
        \centering
        \includegraphics[width=\linewidth]{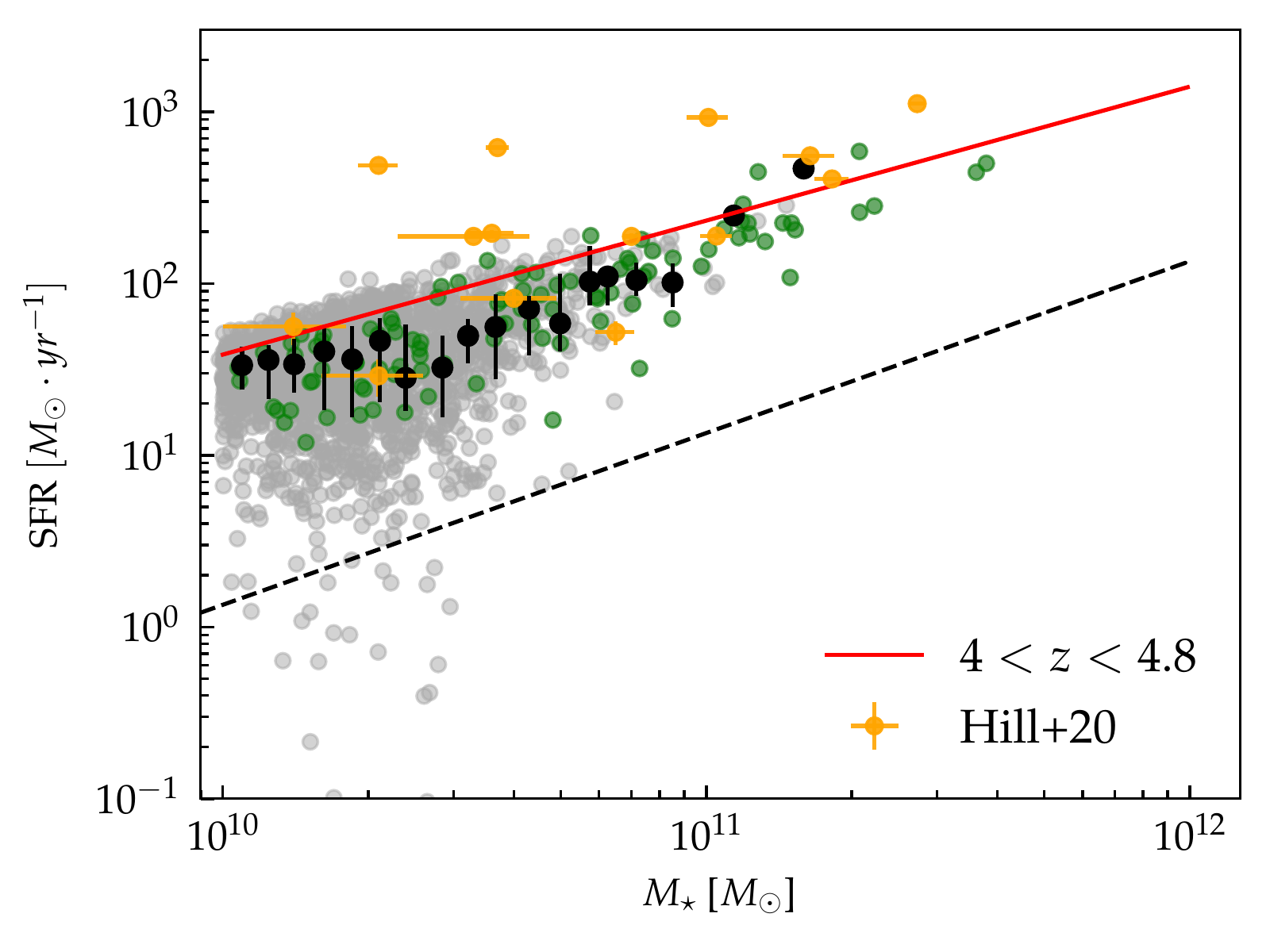}
        \caption{SFR as a function of galaxy stellar mass at $z\sim 4.3$.  Red line are observational data from \cite{2014SEINHARDT}. Orange dots represent galaxies of SPT2349-56 as analysed in \cite{2020HILL}. Grey points are galaxies in our simulations. Black dashed line fix the distinction between quiescent and star forming galaxies (\citealt{2016PACIFICI}). Black points represent median values with 16th ant 84th percentiles for star forming galaxies. Green circles are galaxies from the simulated protoclusters showed in Fig.~\ref{fig:protoz4_sfr}.}
        \label{fig:mstar_sfr_z4}
    \end{figure}
    
   In Fig.~\ref{fig:mstar_sfr_z4} we show the main sequence of star forming galaxies at $4<z<4.8$. The red line represents the main sequence for the field as observed by \cite{2014SEINHARDT}. The orange points are galaxies in SPT2349-56 as reported by \cite{2020HILL}, who updated the values of SFR of \cite{2018MILLER} and estimated the mass of single galaxies by dynamical methods (through their measured line-widths). Grey dots are all galaxies in our simulations at $z=4.3$ with median values and $16^{  \rm th}$-$84^{  \rm th}$ percentiles marked in black. Green points are protocluster galaxies in our simulations. As we can see from the plot, simulations show a fairly good agreement with observations with no statistical difference in the normalisation of the MS. Therefore, we can not explain the difference we observe in terms of SFR by a systematic offset on the SFR-$\rm M_{\star}$ plane. However, if we look at the galaxies in SPT2349-56, we see that they are scattered around the MS with also few strong starburst. On the contrary, galaxies in our simulated protoclusters are mainly MS galaxies with a very small scatter. Therefore, we conclude that at $z\sim 4$ our simulations fail to reproduce the high SFR observed because they are unable to produce strong starburst lying well above the MS. 

 \section{Redshift evolution of mass normalised SFR}
 
\begin{figure}
        \centering
        \includegraphics[width=\linewidth]{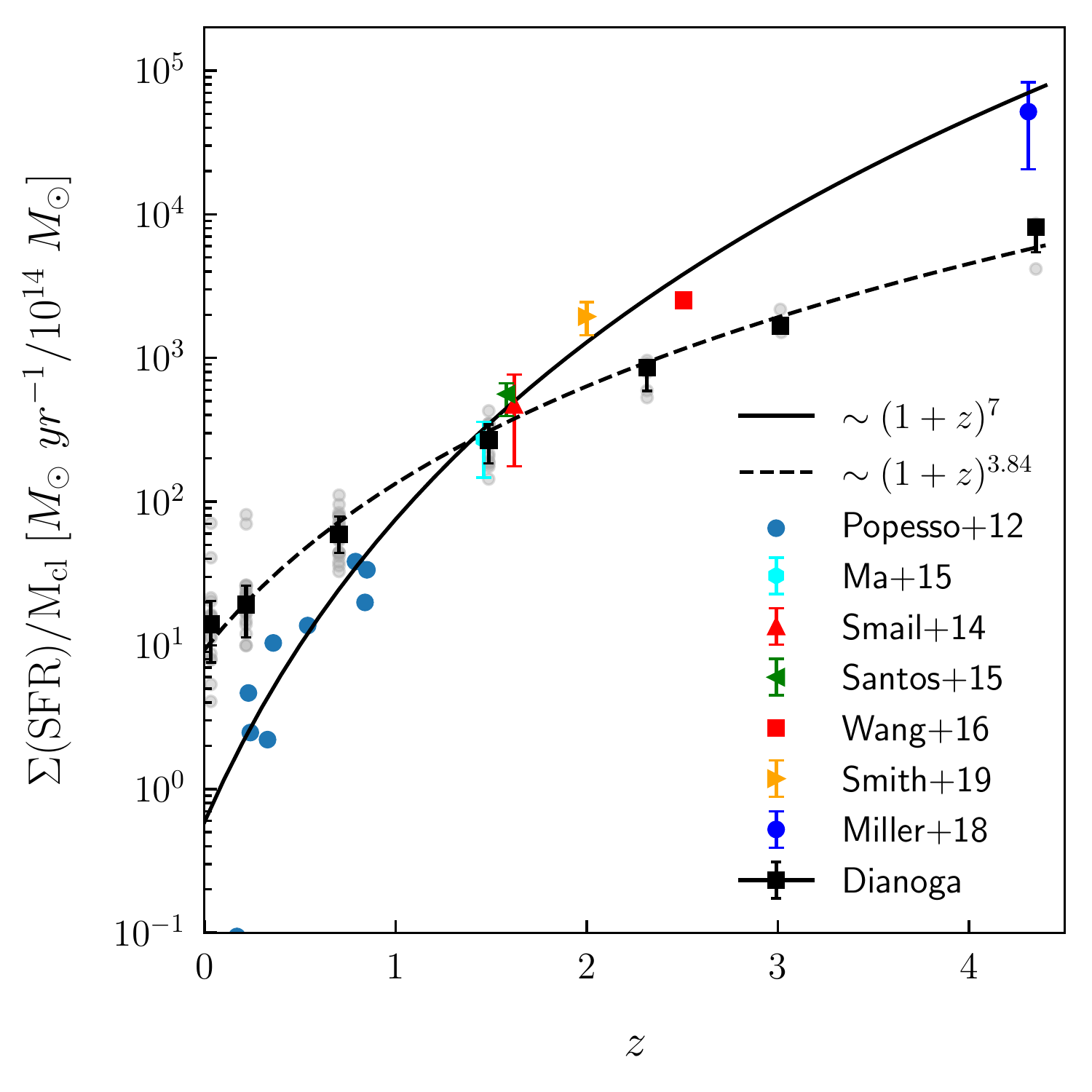}
        \caption{SFR normalised by cluster mass as a function of redshift. Black squares represent median values from Dianoga simulations (grey points). See the text for complete explanation of sample selection. Dashed black line is the best fit to simulations. Coloured points are observational data from \cite{2012POPESSO}, \cite{2015MA}, \cite{2014SMAIL}, \cite{2015SANTOS}, \cite{2016WANG}, \cite{2018MILLER}, and \cite{2019SMITH}. The solid black line $\sim (1+z)^7$ shows an empirical fit to data suggested by \cite{2004COWIE} and \cite{2006GEACH}. }
        \label{fig:mass_sfr}
    \end{figure}

    In this section we show the evolution of the SFR once normalised by the cluster mass, $\Sigma(\rm SFR)/M_{\rm cl}$. The results are showed is Fig.~\ref{fig:mass_sfr}.
 
    %Fig.~\ref{fig:mass_sfr} shows the evolution of the SFR normalized by the cluster mass, $\Sigma(\rm SFR)/M_{\rm cl}$, in simulations and observations.

    $\Sigma(\rm SFR)/M_{\rm cl}$ is an increasing function of redshift with an observationally-driven empirical parametrisation of $(1+z)^n$ (e.g., \citealt{2004COWIE}, \citealt{2006GEACH}).
    \cite{2012POPESSO} used a sample of 9 groups in the redhisft range $0.1 < z < 1.6$ and 9 clusters in the redshift range $0.1 < z < 0.85$ and derived the best fit to be $(213\pm 44)\times z^{1.33 \pm 0.34}$ and $(66\pm 23)\times z^{1.77 \pm 0.36}$ for groups and clusters respectively. \cite{2012POPESSO} also showed that there is no evidence for a significant $\Sigma(\rm SFR)/M_{\rm cl}-M_{\rm cl}$ or ${\rm M}_{\rm cl}-z$ correlation, concluding that $\Sigma(\rm SFR)/M_{\rm cl}-z$ is a genuine correlation (not driven by a decreasing mass evolution with redshift within their sample). Recent observations of highly star forming protocluster regions suggest a stronger evolution with redshift, $\propto (1 + z)^7$ (e.g, \citealt{2014SMAIL}, \citealt{2015MA}, \citealt{2015SANTOS}, \citealt{2019SMITH}), in line with the trend found by \cite{2004COWIE} for the number of star-forming ultraluminous infrared galaxy (ULIRG) in the redshift range $0<z<1.5$. In Fig.~\ref{fig:mass_sfr} we add  the previously cited protocluster regions at high redshift.  We note that for these protoclusters the SFR is computed within an aperture of $\sim 1\ \rm pMpc$, with two exceptions: \cite{2018MILLER} computed the SFR within a 2D aperture of $\sim 130\ \rm pkpc$ and \cite{2016WANG} computed the SFR in a 2D aperture of $\sim 80\ \rm pkpc$. All the SFRs derived assuming the Salpeter IMF are converted to a Chabrier IMF. 
    
    To build the comparison we considered clusters and groups in our simulations with a mass threshold varying with redshift. In particular, the minimum mass considered is $M_{200}>10^{14}$ at $z=0$, decreasing linearly with redshift up to $M_{200}>10^{13}$ at $z=4.3$. We made this choice to mimic as close as possible the minimum mass of the protoclusters and clusters in Fig.~\ref{fig:mass_sfr} at all redshifts. To mark few examples, \cite{2012POPESSO} clusters are in the mass range $[3.9, 27.6]\times 10^{14}\ \rm M_{\odot}$; the cluster by \cite{2019SMITH} at $z\sim2$ has an estimated mass of $0.5\times 10^{14}\ \rm M_{\odot}$, while the protocluster by \cite{2018MILLER} at $z\sim 4.3$ has an estimated mass of $1.16 \pm 0.70\times 10^{13}\ \rm M_{\odot}$. We use $M_{200}$ (like \citealt{2012POPESSO}) as an estimate of the cluster mass and we compute the instantaneous SFR considering all gas particles within $R_{200}$. This choice has the advantage to match the aperture used by \cite{2016WANG} and \cite{2018MILLER}, the two observations at the highest redshifts ($R_{200}\sim 300\ \rm pkpc$ and $R_{200}\sim 150\ \rm pkpc$ at $z=3$ and $z=4$ respectively).

    Simulated clusters show a clear evolution with redshift; however, the trend is shallower than in observations and is better described by $\propto (1+z)^{3.84\pm 0.15}$. In particular, simulations predict a higher SFR at low redshift, reflecting the results already discussed in Fig.~\ref{fig:bcg_sfr}. At redshift $z>2$, the predicted SFR is much lower, mirroring the discussion of the previous Sections. A similar mismatch with respect to observations has also been pointed out by \cite{2018RAGONE} for the sSFR of BCGs of our lower resolution simulations.
%-------------------------------------------------------------------- 

    \begin{figure*}
        \centering
        \includegraphics[width=\linewidth]{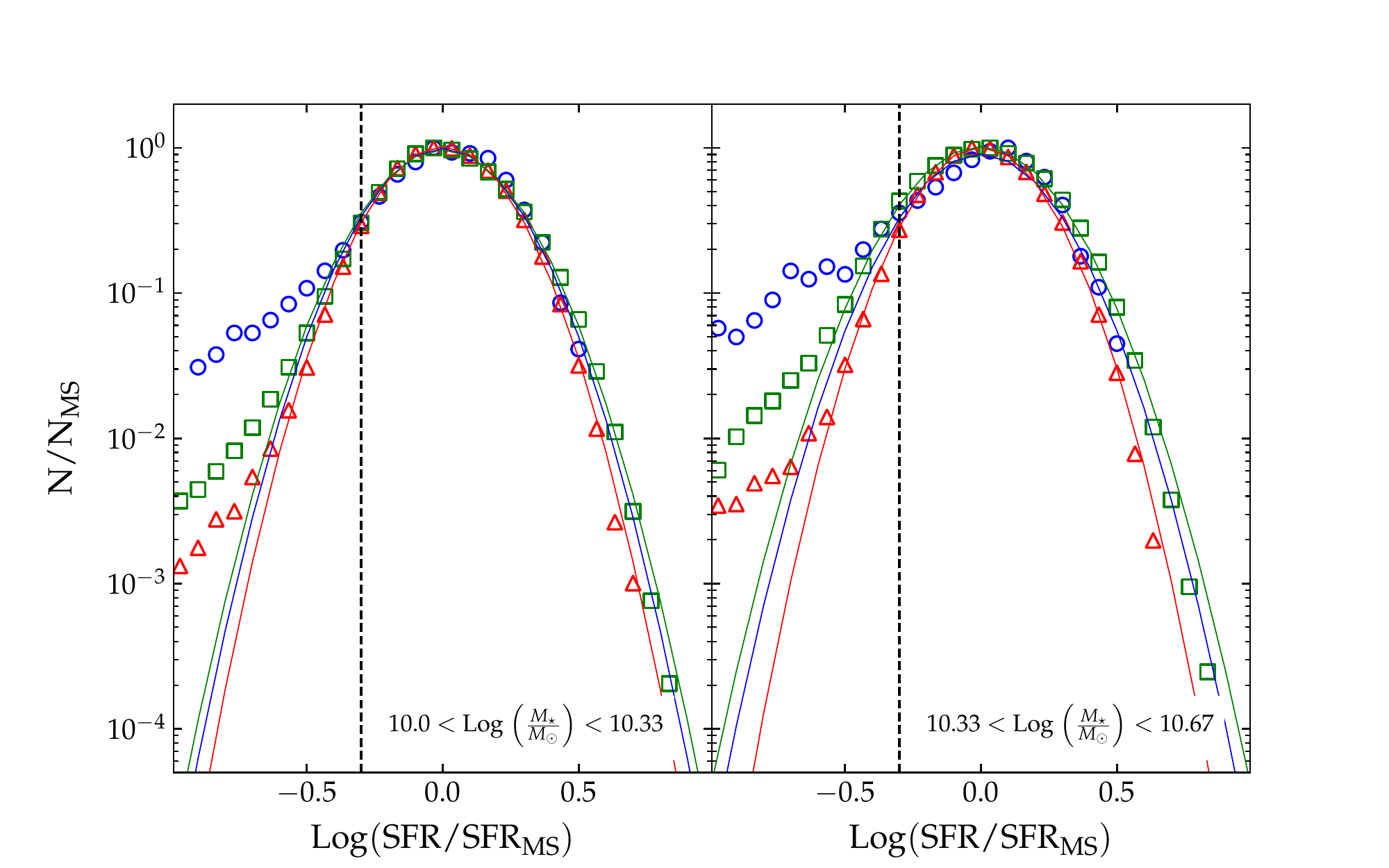}
        \caption{SFR distribution of star forming galaxies at fixed stellar mass at $z=2$. Blue points refer to Dianoga simulations, green squares to Magneticum Box2b and red triangles to Magneticum Box2. $\rm SFR_{\rm MS}$ is computed independently for every simulation. $\rm N_{\rm MS}$ is the number of galaxies within the bin corresponding to SFR=$\rm SFR_{\rm MS}$. Only bins with at least $10$ galaxies are showed. Coloured solid lines are Gaussian fits to simulations. Vertical black dashed line define the threshold above which data are used to estimate the fit.}
        \label{fig:gaussian}
    \end{figure*}
    
\section{Discussion}\label{sec:discussion}

    %but it appears that these galaxies are characterized by both high fraction of molecular gas and a high star formation efficiency (e.g., \citealt{2015BETHERMIN}). Moreover, few authors also suggest that these galaxies are in the stage of galaxy mergers which drive the phase of sturbusting (METTI REFERENZA E SCRIVI MEGLIO). 

    In the recent years a good number of observational studies have confirmed the detection of protocluster regions, characterised by  SFRs from several hundreds to several thousands of ${\rm M}_{\odot}\ \rm yr^{-1}$ (\citealt{2014CLEMENTS}, \citealt{2014DANNERBAUER}, \citealt{2015UMEHATA}, \citealt{2016WANG}, \citealt{2018OTEO}, \citealt{2018COOGAN}, \citealt{2018MILLER}, \citealt{2019GOMEZ}, \citealt{2019SMITH}, \citealt{2019LACAILLE}). These high values of SFR are often dominated by DSFGs, with typical SFRs from $\sim 100\ \rm M_{\odot}\ \rm yr^{-1}$ to $\sim 1000\ \rm M_{\odot}\ \rm yr^{-1}$. The physical reason of these high values of SFR is not settled. Some observations suggest that starburst galaxies and SMGs are characterised by a high star formation efficiency (e.g., \citealt{2010DADDI}). Other recent observations suggest that the starbursting phase of galaxies is related to high gas fractions (e.g., \citealt{2016SCOVILLE},
    \citealt{2019GOMEZ}). Finally, some observations suggest that starburst galaxies are characterised by both high star formation efficiency and high gas fraction (\citealt{2015GENZEL}, \citealt{2015BETHERMIN}).
    
    In this section we focus on starburst galaxies in our simulations and the differences in terms of star formation efficiency and gas fraction with respect to galaxies in the observed protoclusters used as references in Sect.~\ref{sec:proto_z2} and Sect.~\ref{sec:proto_z4}. We also investigate the implications for the subresolution model of star formation adopted in our simulations.
    
    \subsection{Starburst galaxies in numerical simulations}\label{sec:discussion_sb}
    
    \begin{figure}
        \centering
        \includegraphics[width=\linewidth]{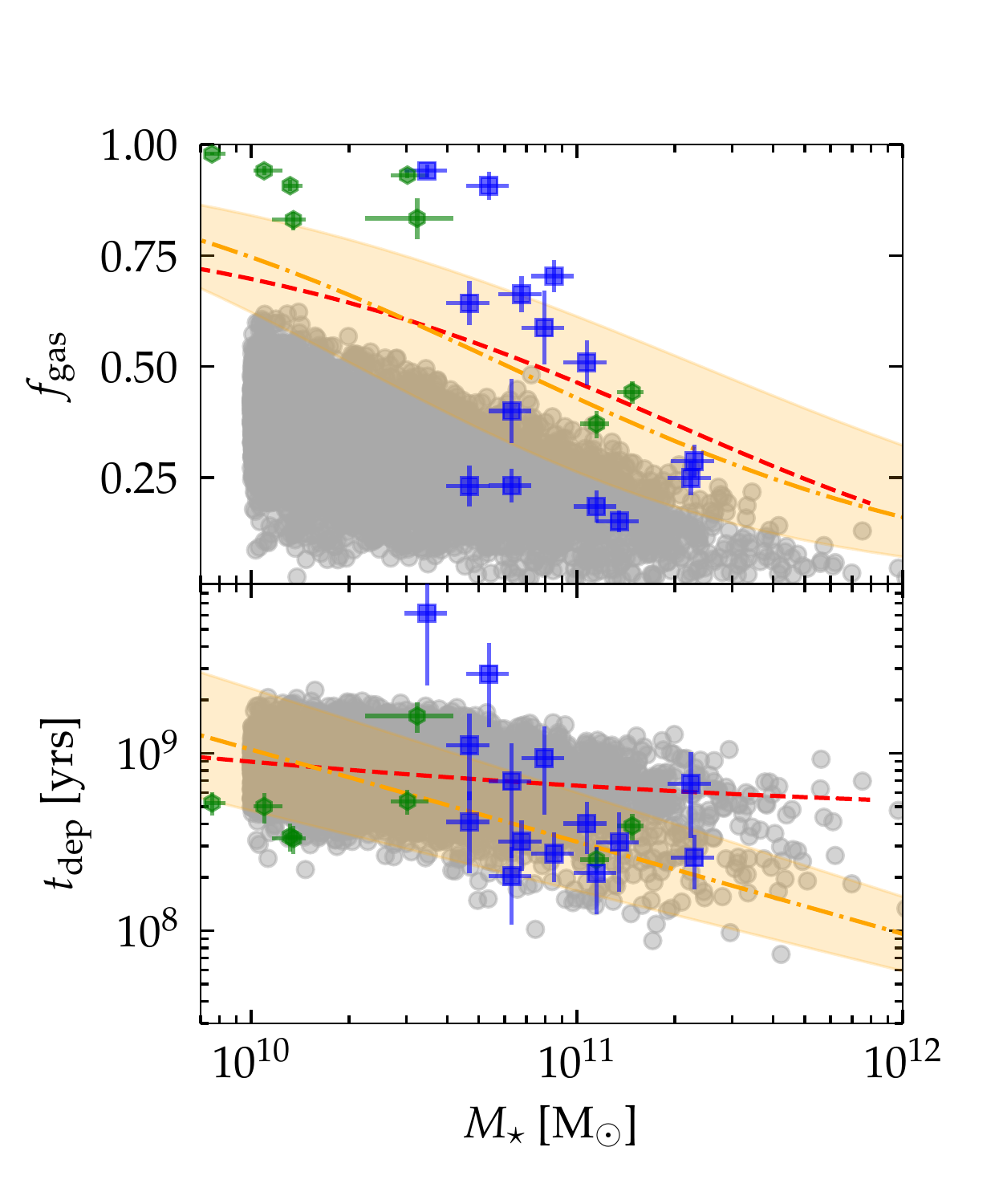}  
        \caption{Galaxy correlations at $z=2.3$. {\it Top panel:} gas fraction as a function of stellar mass. {\it Bottom panel:} depletion time as a function of stellar mass. Grey circles refer to Dianoga simulations at z=2.3. Green hexagons and blue squares are data from \cite{2019GOMEZ} and \cite{2018WANG} respectively. Orange dashed line is the functional form of \cite{2019LIU} for main sequence galaxies at $z=2.3$, while the shaded region encompasses galaxies with SFR 4 times lower and higher than main sequence galaxies. Red dashed lines are obtained combining the MS by \cite{2014WHITAKER} and the integrated Kennicutt-Schmidt law from \cite{2014SARGENT} (see text for further details).}
        \label{fig:dept_z2}
    \end{figure}
    
    Galaxies are usually defined as starburst depending how much their SFR is above the SFR of main sequence galaxies with the same mass and at the same redshift. Here, following \cite{2015SCHREIBER}, we consider the threshold SFR/$\rm SFR_{\rm MS} > 4$. In the $M_{\star}-$SFR plane, starburst galaxies does not only represent the tail of the MS distribution. Indeed several studies showed that at fixed stellar mass and redshift, the distribution of galaxies around the main sequence is better described by a double Gaussian (\citealt{2012SARGENT}, \citealt{2015SCHREIBER}), where the second component describes the population of starburst galaxies. This population is estimated to comprise $3\%$ of star forming galaxies without significant redshift dependence (\citealt{2015SCHREIBER}).

    Following the works mentioned above we study the starburst population in our simulations, by plotting the distribution of galaxies around the main sequence in two mass bins at $z=2$, see Fig.~\ref{fig:gaussian}. Since in principle it is not guaranteed that the most star forming galaxies will be within protocluster regions, we also plot the results from the Magneticum simulations\footnote{http://www.magneticum.org/}. The Magneticum simulations are a set of hydrodynamical simulations of different cosmological volumes (\citealt{2014HIRSCHMANN}, \citealt{2017RAGAGNIN}), performed with the same GADGET-3 code used in our simulations (see \citealt{2014HIRSCHMANN} for the differences in the AGN feedback implementation). From the Magneticum set, we consider the Box2 and Box2b (352 and 640 $h^{-1}\ \rm Mpc$ respectively). The mass resolution is $m_{\rm DM}=6.9\times 10^{8}\ h^{-1}\ M_{\odot}$, a factor of $10$ lower than the one used for this work.
    
    %Since the volume sampled in our simulations is small, we also plot the results from the Magneticum simulations\footnote{http://www.magneticum.org/}. The Magneticum simulations are a set of hydrodynamical simulations of different cosmological volumes (\citealt{2014HIRSCHMANN}, \citealt{2017RAGAGNIN}), performed with the same GADGET-3 code employed in our simulations (see \citealt{2014HIRSCHMANN} for the differences in the AGN feedback implementation). From the Magneticum set, we consider the Box2 and Box2b (352 and 640 $h^{-1}\ \rm Mpc$ respectively). The mass resolution is $m_{\rm DM}=6.9\times 10^{8}\ h^{-1}\ M_{\odot}$, a factor of $10$ lower than the one used for this work. Moreover, in addition to the increased statistics, it is also useful to compare with cosmological boxes as . 
    
    \begin{figure}
        \centering
        \includegraphics[width=\linewidth]{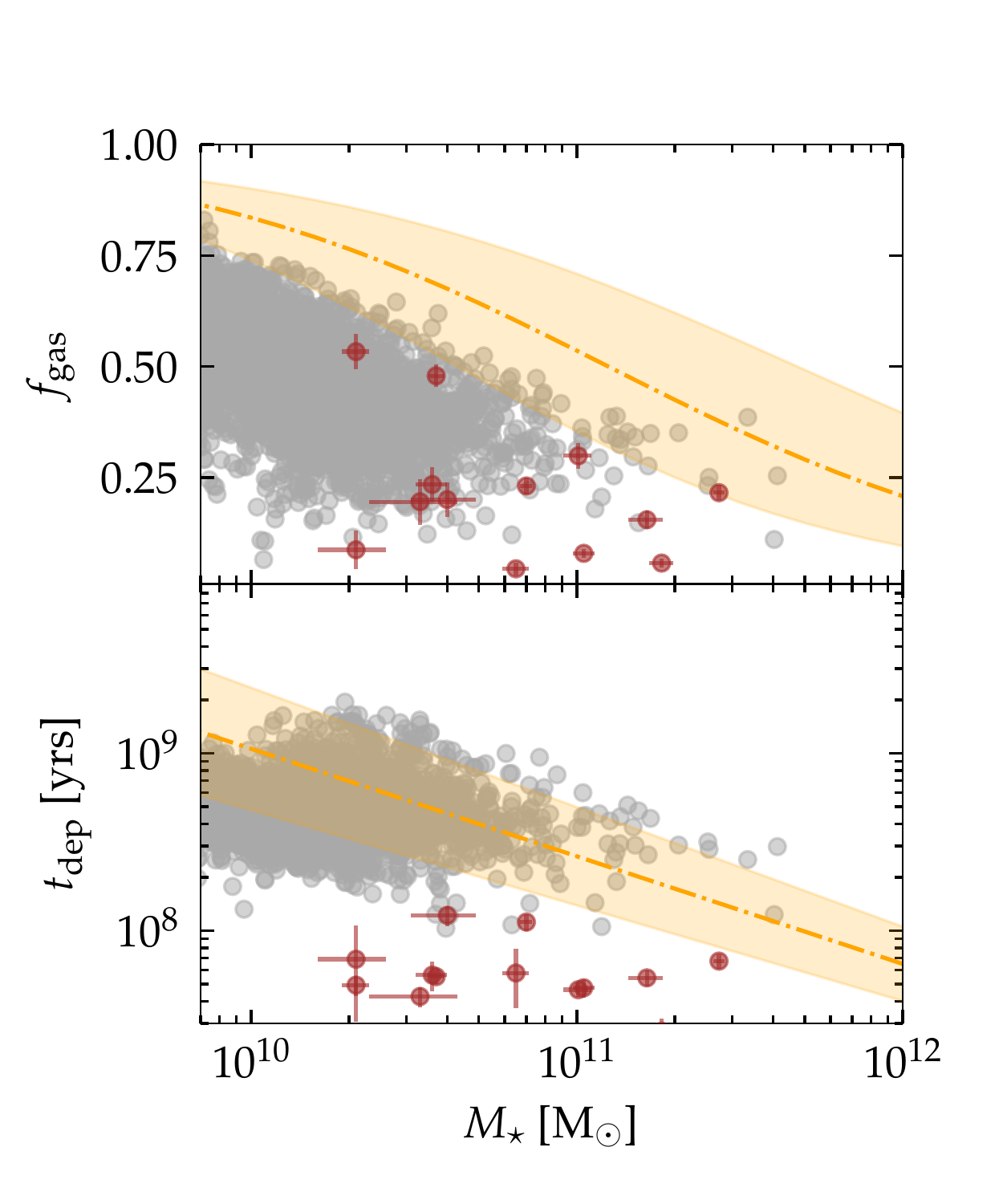}
        \caption{Galaxy correlations at $z=4.3$. {\it Top panel:} gas fraction as a function of stellar mass. {\it Bottom panel:} depletion time as a function of stellar mass. Grey circles refer to Dianoga simulations at z=4.3. Brown circles are data from \cite{2020HILL}. Orange dashed line is the functional form of \cite{2019LIU} for main sequence galaxies at $z=4.3$, while the shaded region encompass galaxies with SFR 4 times lower and higher than main sequence galaxies. }
        \label{fig:dept_z4}
    \end{figure}

    \begin{figure}
        \centering
        \includegraphics[width=\linewidth]{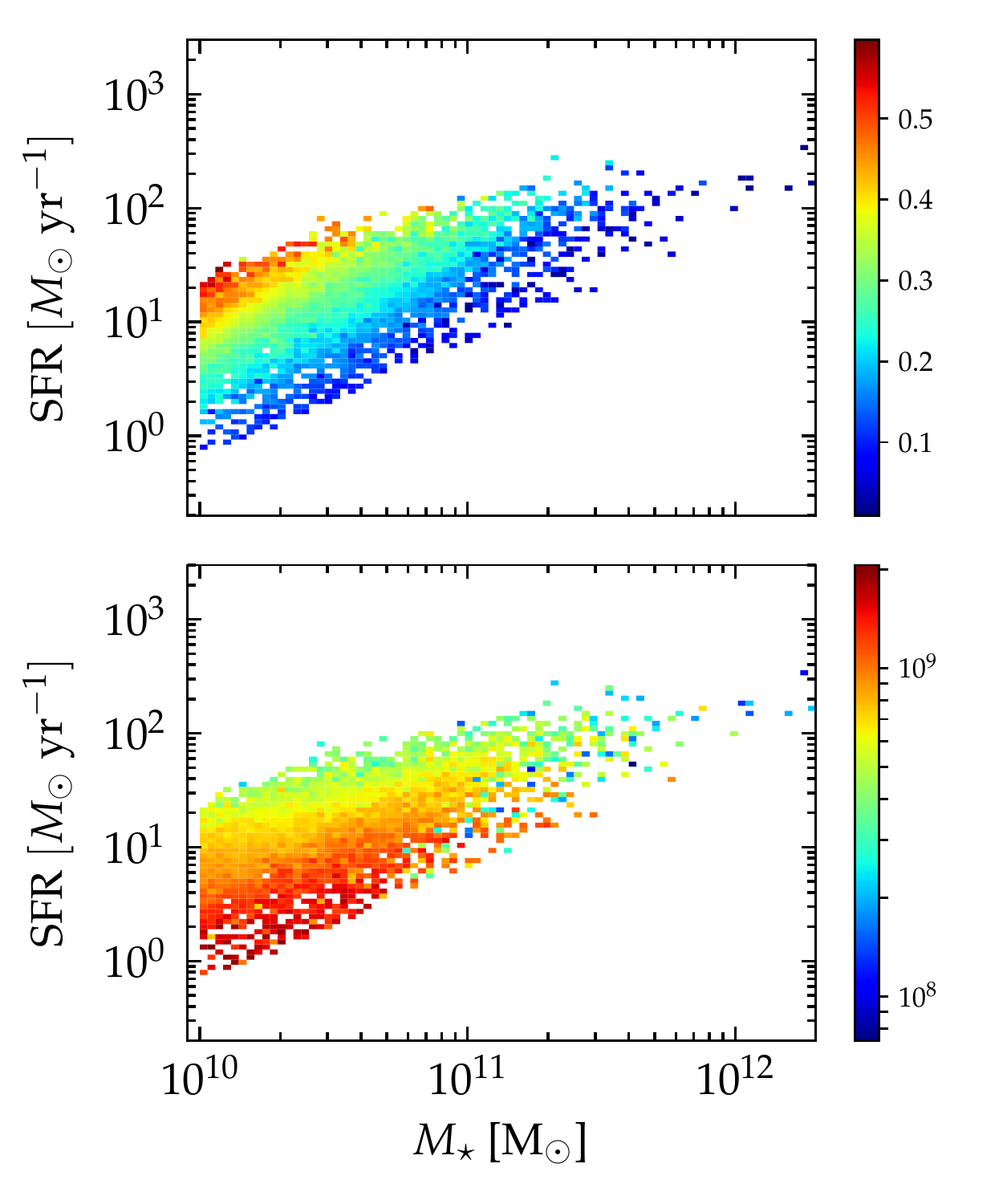}
        \caption{{\it Top panel:} 2D histogram of main sequence star forming galaxies in simulations at $z=2.3$. Each bin is colour-coded with the respective median value of $f_{\rm gas}$ {\it Bottom panel:}  Same as upper panel, colour-coded with respect to $t_{\rm dep}$ }
        \label{fig:ms_colorcoded}
    \end{figure}
    
    In Fig.~\ref{fig:gaussian} the value of $\rm SFR_{\rm MS}$ is computed for each simulation considering only active galaxies (see Sect.~\ref{sec:proto_z2}, Fig.~\ref{fig:mstar_sfr}). The two mass bins analysed are chosen following \cite{2012SARGENT}. We analysed only the two lower mass bins as at higher masses the number of galaxies in the Dianoga simulations is too low for a statistical analysis. Blue points are Dianoga simulations, red triangles are results for Box2 and green squares results for Box2b. Only bins with at least ten galaxies are plotted. Solid lines are Gaussian fits to the data, obtained considering only galaxies with $\rm SFR > 0.5\times SFR_{\rm MS}$. This cut, marked as a vertical black dashed line in the plot, is needed to avoid that galaxies on their way to be quenched (but still selected as active by our cut in sSFR) take effect to the Gaussian fit. The three simulations are in very good agreement in both mass bins, despite the different box sizes and environment, and can all be fitted with a single Gaussian with a standard deviation estimated to be $0.19 < \sigma < 0.21$. This value is in agreement with the results of \cite{2012SARGENT} ($\sigma = 0.188$), but lower than the estimate of \cite{2015SCHREIBER} ($\sigma = 0.31$). Finally, the fraction of starburst galaxies (i.e., SFR/$\rm SFR_{\rm MS} > 4$ ) is $0.03\% < f_{\rm SB} < 0.2\%$, at least one order of magnitude lower than what estimated by \cite{2015SCHREIBER}.
    
    As a final warning, it is also important to keep in mind that the observational estimates of the values of SFR, in particular for starburst galaxies, are affected by a number of uncertainties. A relevant role is played by the assumption of the IMF, as studies on the chemical abundances and abundance ratios suggest that starburst galaxies are characterised by a top-heavy stellar IMF (e.g., \citealt{2017ROMANO}, \citealt{2019ROMANO}). 
    %SB. Dire in che direzione cambierebbero i risultati.
    If it is proved right, this could have an important implication on the estimated values of SFR, affecting as a consequence also the conclusions reached in this sub-section. Indeed, a more top-heavy IMF would lower the SFR values obtained through observations, while affecting much less numerical predictions (see also the discussion by \citealt{2015GRANATO}).
    
    \begin{figure*}
        \centering
        \includegraphics[width=\linewidth]{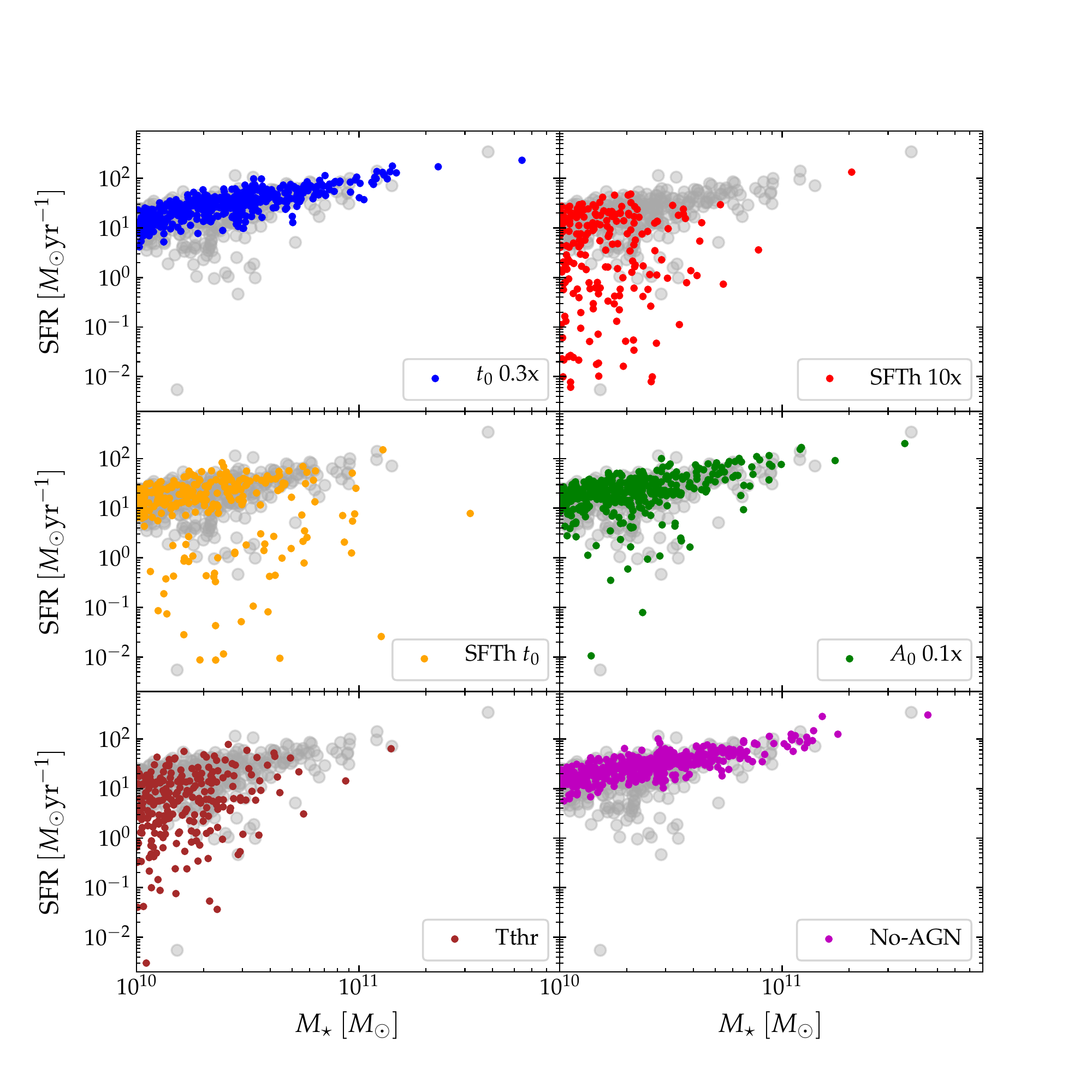}
        \caption{Main sequence of star forming galaxies at $z=3$ for different simulations. Grey points refer to the results relative to the same region used for the tests with the set up used for this work. Different panels refer to: {\it $t_0\ 0.3$x}: shorter time-scale for star formation; {\it SFTh $10$x}: increased density threshold  for star formation; {\it SFTh} $t_0$: increased density threshold for star formation and shorter star formation time-scale; {\it $A_0\ 0.1$x}: reduced supernovae thermal feedback; {\it Tthr}: AGN feedback implementation as in \cite{2018RAGONE}; {\it No-AGN}: no AGN feedback. }
        \label{fig:MS_comparison_z3}
    \end{figure*}
    
    \begin{figure*}
        \centering
        \includegraphics[width=\linewidth]{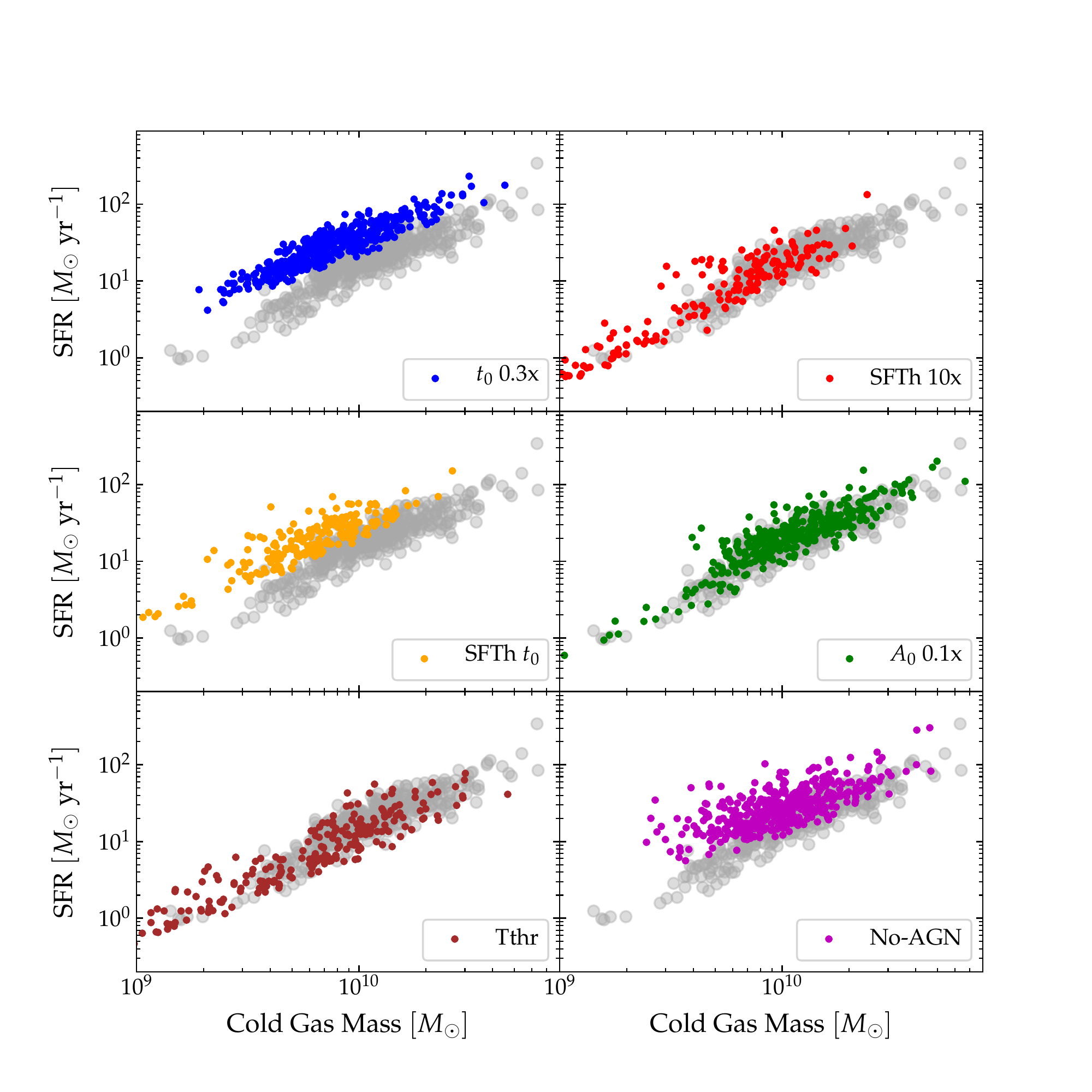}
        \caption{Correlation between cold gas mass and SFR at $z=3$ for different simulations. Grey points refer to the results relative to the same region used for the tests with the set up used for this work. Different panels refer to: {\it $t_0\ 0.3$x}: shorter time-scale for star formation; {\it SFTh $10$x}: increased density threshold  for star formation; {\it SFTh} $t_0$: increased density threshold for star formation and shorter star formation time-scale; {\it $A_0\ 0.1$x}: reduced supernovae thermal feedback; {\it Tthr}: AGN feedback implementation as in \cite{2018RAGONE}; {\it No-AGN}: no AGN feedback. }
        \label{fig:SFE_comparison_z3}
    \end{figure*}
    
    \begin{figure*}
        \centering
        \includegraphics[width=\linewidth]{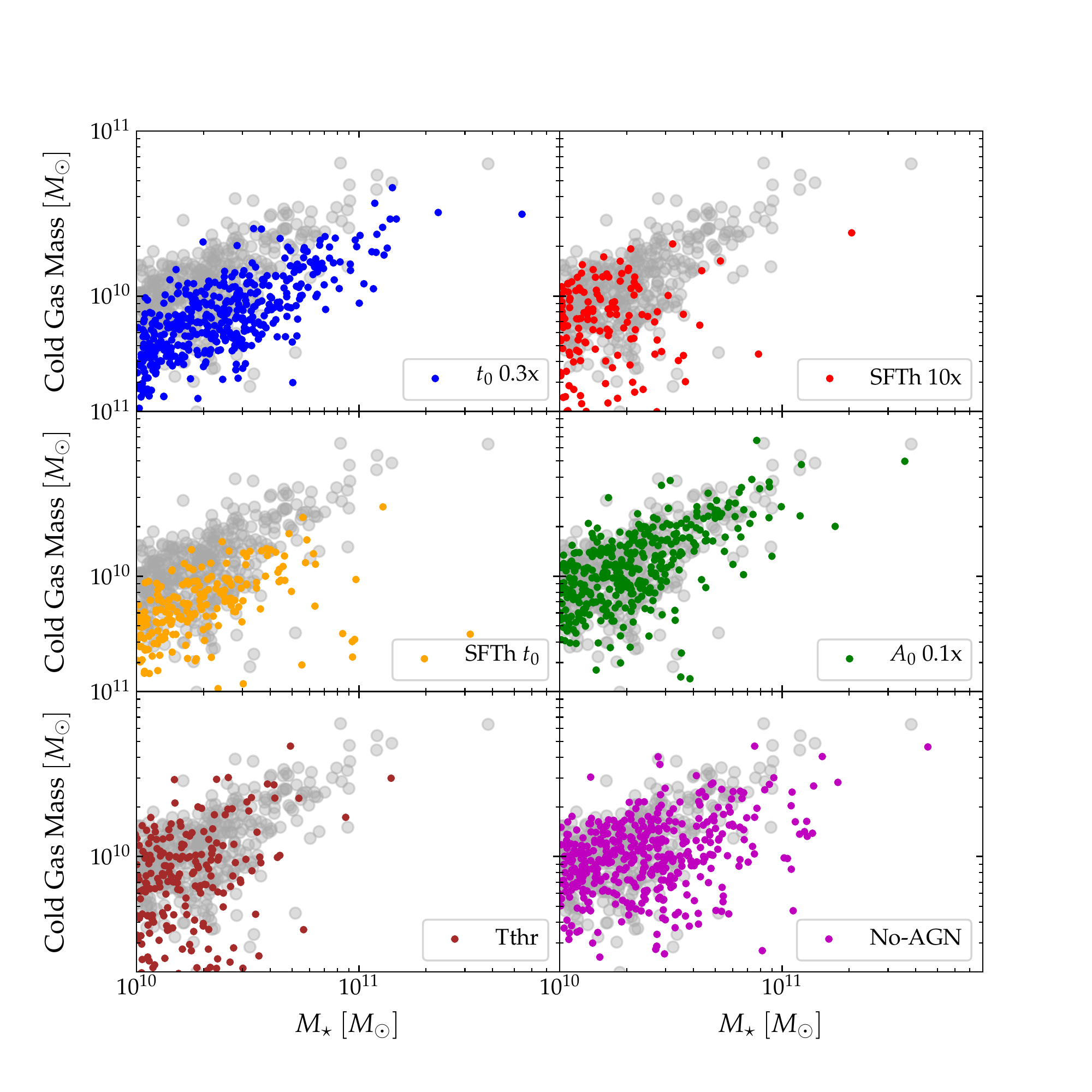}
        \caption{Correlation between stellar mass and cold gas mass at $z=3$ for different simulations. Grey points refer to the results relative to the same region used for the tests with the set up used for this work. Different panels refer to: {\it $t_0\ 0.3$x}: shorter time-scale for star formation; {\it SFTh $10$x}: increased density threshold  for star formation; {\it SFTh} $t_0$: increased density threshold for star formation and shorter star formation time-scale; {\it $A_0\ 0.1$x}: reduced supernovae thermal feedback; {\it Tthr}: AGN feedback implementation as in \cite{2018RAGONE}; {\it No-AGN}: no AGN feedback. }
        \label{fig:gas_fraction_comparison_z3}
    \end{figure*}

    \subsection{Star formation efficiency and gas fraction}\label{sec:discussion_SFE}
    
    In this Section we study the cold gas and SFR properties of our simulated galaxies, to determine which variable is more related to the underestimated normalisation of the main sequence at $z\sim 2$ (see Sect.~\ref{sec:proto_z2}) and to the absence of starburst galaxies.
    
    In Fig.~\ref{fig:dept_z2} and Fig.~\ref{fig:dept_z4} we study the gas fraction and the depletion time in simulations and observations at $z\sim 2$ and $z\sim 4$ respectively. The two are defined as
    \begin{equation}
        f_{\rm gas} = \frac{M_{\rm gas}}{M_{\rm gas} + M_{\star}},
    \end{equation}
    where $M_{\rm gas}$ is the cold gas mass that in simulations is computed considering only the cold phase of SPH particles, and 
    \begin{equation}
        t_{\rm dep} = \frac{M_{\rm gas}}{\rm SFR}.
    \end{equation}
    We plot at $z\sim2$ and at $z\sim4$ all the galaxies in the observed protoclusters used in Sect.~\ref{sec:proto_z2} and Sect.~\ref{sec:proto_z4} respectively (green hexagons, blue squares and orange circles represent galaxies in \citealt{2019GOMEZ}, \citealt{2018WANG}, and \citealt{2020HILL}). We plot as orange dashed line the functional forms of $f_{\rm gas}$ and $t_{\rm dep}$ from \cite{2019LIU}. The functional forms depend on redshift, stellar mass and relative distance from the main sequence ($\rm SFR/SFR_{\rm MS}$). The orange dashed line refers to main sequence galaxies, while the shaded region encompasses galaxies with a SFR 4 times lower and higher than MS galaxies.  The red dashed line in Fig.~\ref{fig:dept_z2} is computed as follow: given a galaxy stellar mass, we assume that the expected SFR for a main sequence galaxy is given by the MS relation of \cite{2014WHITAKER}. Given the SFR, we assume that the mass of molecular gas is given by eq. $4$ of \cite{2014SARGENT} for normal galaxies. Combining $M_{\star}$, $M_{\rm gas}$, and SFR we obtain the expected values of $f_{\rm gas}$ and $t_{\rm dep}$ for normal star forming galaxies. This procedure is supported also by recent observations with ALMA, which show that massive ($M_{\star}>10^{10}\ M_{\odot}$) main sequence galaxies obey the star-forming galaxies’ star formation law (\citealt{2019LIU}). From Fig.~\ref{fig:dept_z2} and Fig.~\ref{fig:dept_z4} we see that observational data are not in agreement among each other. In particular, data from \cite{2020HILL} at $z\sim 4$ suggests that starburst galaxies are characterised by a short depletion time (and thus a high star formation efficiency) than normal star forming galaxies. This is supported also by other observations (i.e., \citealt{2010DADDI}, \citealt{2014SARGENT}, \citealt{2019LIU}). On the other hand, data from \cite{2019GOMEZ} and \cite{2018WANG} at $z \sim 2.4$ are characterised by a high gas fraction but a star formation efficiency (or depletion time) consistent with the integrated Kennicutt-Schmidt law for normal star forming galaxies. This is also consistent with numerical simulations, where the position on the main sequence depends on the gas fraction (see Fig.~8 of \citealt{2019DAVE}). The difference among the observational results is largely due to the different values of the parameter $\alpha_{\rm CO}$  used to derive the molecular gas mass from the CO line luminosity $L'_{CO(1-0)}$ through Eq.~\ref{eq:CO}. \cite{2019GOMEZ} used $\alpha_{\rm CO}=3.5$, typical for normal star forming galaxies at solar metallicity.  \cite{2018WANG} adopts the mass-metallicity relation by \cite{2015GENZEL} to retrieve the galaxy metallicity for the members of their cluster, and then computed the metalicity dependent value of $\alpha_{\rm CO}$ following \cite{2015GENZEL} and \cite{2018TACCONI}. The values reported for the $\alpha_{\rm CO}$ are in the range $[4.06, 4.12]$, again consistent with normal star forming galaxies at solar metallicity. \cite{2020HILL}, on the other hand, used $\alpha_{\rm CO}=1$, typical for high redshift SMGs. Due to these arguments, it remains unclear whether starburst galaxies in protocluster environment are mainly driven by a high gas fraction or a high star formation efficiency, as the results strongly depend on the assumptions needed to derive gas related quantities.

    If we look at the population of normal star forming galaxies at $z\sim 2$ (red and orange dashed lines in Fig.~\ref{fig:dept_z2}) we see that our simulations match the observed star formation efficiency. On the other side, the simulated gas fractions are consistently lower than observations. In particular, at $M_{\star} = 10^{10}\ M_{\odot}$, observations have a $M_{\rm gas}$ higher by a factor of $\sim 3.5$. This factor is very similar to the difference in the observed and simulated main sequence (see Fig.~\ref{fig:mstar_sfr}). Thus, the lower normalisation in the simulated main sequence seems to be driven by an underestimated gas fraction.
    
    It is also interesting to study the correlation between the main sequence galaxies and gas-related properties in simulations. In Fig.~\ref{fig:ms_colorcoded} we show the MS colour-coded with respect to $f_{\rm gas}$ (upper panel) and $t_{\rm dep}$ (lower panel). At fixed stellar mass, galaxies below the MS are characterised by both a long depletion time and low gas fraction. Vice-versa, galaxies above the MS have both short depletion times and high gas fraction. This visual impression is also confirmed by the computation of the Pearson's correlation coefficient for the two relations: $\rm Log(SFR/SFR_{\rm MS})-\rm Log(f_{\rm gas})$ and $\rm Log(SFR/SFR_{\rm MS})-\rm Log(t_{\rm dep})$. The results are $r=0.62$ and $r=-0.63$ for the two relations respectively. Moreover, results of the two linear regressions suggest that in our simulations the position on the main sequence scales with $f_{\rm gas}$ and $t_{\rm dep}$ with similar slopes: SFR/SFR$_{\rm MS}\propto f_{\rm gas}^{0.85}$ and SFR/SFR$_{\rm MS}\propto t_{\rm dep}^{-1}$.

    \subsection{Simulation tests}\label{sec:tests}
    
    In the previous sections we showed that our simulations underestimate the normalisation of the main sequence relation at $z\sim 2$ by a factor of $\sim 3$. Moreover, we have seen that the star formation model (\citealt{2003SPRINGEL}) implemented in our code, with the current choice for the model parameters set to reproduce quiescent mode of star formation, does not reproduce the observed population of starburst galaxies at $z>2$. While the normalisation of the MS seems to be mainly related to the gas fraction, it remains unclear whether we miss starburst galaxies because we do not correctly sample the star formation efficiency, the gas fraction, or both. Therefore, we performed a set of simulations aiming at checking whether the fraction of starburst galaxies and MS normalisation are sensitive to the choice of the parameters of the subgrid model. All the following simulations are performed for only one of our regions, a cluster with $M_{200}= 5.4\times 10^{14}\ M_{\odot}$ at $z=0$. In Fig.~\ref{fig:MS_comparison_z3}, Fig.~\ref{fig:SFE_comparison_z3}, and Fig.~\ref{fig:gas_fraction_comparison_z3} we show the results for the main sequence, SFE, and gas fraction at $z\sim 3$. The choice of the redshift is somewhat arbitrary, as we do not aim at comparing simulations with particular observational data but to study the effect of different parameters on our results. Each panel refers to a different simulation, while we plot with grey circles the results for the reference simulation used in the previous sections. In the following we briefly discuss the specific changes for each simulation and the effects on the results.

    \subsubsection{Increasing the  star formation efficiency ($t_0\ 0.3$x)}
    
    We recall that in \cite{2003SPRINGEL} model the characteristic time for the star formation, $t_{\star}$, is $t_{\star}\propto t_0^{\star}\ t_{\rm dyn}$ where $t_0^{\star}$ is a parameter usually tuned to reproduce the Kennicutt relation (see Sect.~\ref{sec:effective_model}). Here we increase the efficiency to match the observed SFE of \cite{2020HILL} (see Fig.~\ref{fig:dept_z4}) by lowering $t_0^{\star}$ by a factor of 3. The results of this test are shown in the top-left panel of the figures. From Fig.~\ref{fig:SFE_comparison_z3} we see that indeed the SFE is higher, but there is little difference in the main sequence (see Fig.~\ref{fig:MS_comparison_z3}). Moreover, there is no difference in the fraction of starburst galaxies. In fact, the model is so tightly self-regulated that in response to a high SFE we have a lower gas fraction (see Fig.~\ref{fig:gas_fraction_comparison_z3}), resulting in similar SFRs. 
    
    \subsubsection{Increasing the star formation threshold ($\rm SFTh\ 10x$ \& $\rm SFTh\ t_0$)}
    
    In the $\rm SFTh\ 10x$ simulation we increased by a factor of 10 the density threshold, $\rho_{\rm thr}$, used to decide whether a gas particle becomes multiphase (we recall that only multiphase particles can form stars, see Sect.~\ref{sec:effective_model} and \citealt{2003SPRINGEL}). Increasing this threshold should allow to accumulate larger reservoir of gas and reach higher densities before starting to produce stars, increasing the gas fraction and the overall SFR. However, from the top-right panel of Fig.~\ref{fig:MS_comparison_z3}, Fig.~\ref{fig:SFE_comparison_z3}, and Fig.~\ref{fig:gas_fraction_comparison_z3} we see that we do not have major differences in terms of MS normalisation, SFE and gas fraction. The only appreciable difference is the reduction of the most massive galaxies and the increase of passive galaxies. indeed, higher densities at the centre of galaxies also mean more gas accretion onto the central BH and consequently a stronger AGN feedback. 
    
    In the $\rm SFTh\ t_0$ run (central-left panel) we both increased the density threshold for multiphase particles by a factor of 10 and the SFE by a factor of 3. Again, the self-regulation of the star formation model and the AGN feedback prevent any appreciable difference with respect to our fiducial run.

    \subsubsection{Increasing time-scale for cold gas evaporation ($A_0\ 0.1\rm x$)}
    
    Following \cite{2003SPRINGEL}, even if the subgrid model is explicitly constructed to reproduce quiescent star formation, starburst should arise whenever the timescale for star formation is shorter than the timescale for the evaporation of cold gas. In fact, in this regime self-regulation is expected to break down with cold gas transformed into stars before it can be evaporated by stellar feedback. In practice, the relation that should be satisfied is: 
    \begin{equation}\label{eq:sb_cond}
        \frac{t_c}{t_{\star}} = \left( \frac{\rho}{\rho_{\rm thr}} \right)^{4/5} \frac{1}{\beta A_0} > 1 ,
    \end{equation}
    where $\rho_{\rm thr}$ is the density threshold for a particle to become multiphase, $\beta$ is the fraction of stars that instantly die as supernovae, and $A_0$ is a parameter of the model that defines the energy of supernovae used to evaporate cold gas. In this test we reduced the value of $A_0$ by a factor of 10. From the results shown in the central-right panels we see that we do not have any improvement in terms of starburst galaxies. Thus, even if we checked that single gas particles satisfy Eq.~\ref{eq:sb_cond}, this is not sufficient to have a high enough integrated value of SFR. 
    
    \subsubsection{Varying AGN feedback implementation (Tthr)}
    
    To quantify the effect of a specific aspect of the  AGN feedback implementation on our results we also run a simulation  with the same AGN feedback prescription of \cite{2018RAGONE}. We recall that in that set-up there is an extra condition on the temperature (T<T$_{\rm thr}$) to consider a particle as multiphase and that the energy released by AGN feedback is used to evaporate molecular clouds, while in the current implementation is coupled only to the hot phase of multiphase particles. From the bottom-left panels of Fig.~\ref{fig:MS_comparison_z3}, Fig.~\ref{fig:SFE_comparison_z3}, and Fig.~\ref{fig:gas_fraction_comparison_z3} we can see that the only difference with respect to our fiducial run is that in this case we have less massive galaxies. This is expected from the results showed in Sect.~\ref{sec:gsmf} and Sect.~\ref{sec:mstar-Mbcg}, where it was clear that the feedback implementation of \cite{2018RAGONE} is more effective in quenching star formation. 
    
    \subsubsection{No AGN feedback (No-AGN)}
    
    Finally, we also performed a simulation without AGN feedback (bottom-right panels). This is of course to test an extreme scenario, as the absence of AGN feedback would result in GSMF, BCG masses and SFR inconsistent with low-redshift observations. From Fig.~\ref{fig:MS_comparison_z3}, we see that in this run we have fewer galaxies on their way to become passive and more massive galaxies, as expected. However, the MS retain the same normalisation and there is no signature for an increased fraction of starburst galaxies. Moreover, it is interesting to note that in the {\it No-AGN} run the SFE is higher in the low mass regime (see Fig.~\ref{fig:SFE_comparison_z3}). This difference is due to the fact that without AGN feedback the gas reaches higher density, especially in the low mass regime where the feedback is more efficient in expelling gas outside the shallow potential wells of galaxies. 
    
    \section{Conclusions}
    In this paper we studied the SFR of simulated protocluster regions and the gas properties of protoclusters galaxies in the redshift range $2 < z < 4$, and we compared them with observations. Our work is based on a subsample of the Dianoga simulations (\citealt{2011BONAFEDE}). In particular, we used 12 clusters, 7 of which very massive ($M_{200} > 8 \times 10^{14}h^{-1}\ M_{\odot}$). The simulations are carried out with GADGET3, a modified version of the public code GADGET2, which implements a SPH scheme for hydrodynamics and treats the unresolved baryonic physics through various subgrid models. In particular, we use the \cite{2003SPRINGEL} model for star formation and a thermal AGN feedback. In Sect.~\ref{sec:gsmf} and Sect.~\ref{sec:mstar-Mbcg} we presented the degrees of freedom of the AGN feedback implementation. With the implementation of \cite{2018RAGONE}, where a temperature threshold is used to define multiphase gas particles and the energy released by AGN feedback is used to evaporate their cold phase, we match the observed correlation between cluster and BCG mass, but the normalisation of the galaxy stellar mass function is lower by a factor of $\sim 2$ with respect to observations (see \citealt{2019BASSINI}, Appendix B). On the other hand, without the temperature threshold and coupling the energy released by AGN feedback only with the hot phase of gas particles, we match the GSMF but we get too massive BCGs (a factor of $\sim2$, see Sect.~\ref{sec:mstar-Mbcg}). In this paper, we use the latter implementation, which maximises the value of the SFR in the high-redshift regime ($z \sim 2-4$) in which we are interested. Our main results can be summarised as follow:
    
    \begin{itemize}
        \item At $z \sim 2$ simulations under-predict the SFR of highly star forming protocluster regions by a factor of 4 or even larger, in line with the results we presented in \cite{2015GRANATO}, based on a larger set of lower resolution simulations. This result is indeed stable against numerical resolution and is the combination of two effects: $(i)$ simulations under-predict the normalisation of the main sequence at $2<z<2.5$ by a factor of 3; $(ii)$ simulations predict a fraction of starburst galaxies, defined as galaxies with a SFR at least four times higher than main sequence galaxies, of $[0.2\% - 0.03\%]$, at least a factor of ten lower than what recent observations find (\citealt{2015SCHREIBER}). We verified that this result is independent of the environment 
        by performing the same analysis on the Magneticum cosmological boxes of 352 and 640 $h^{-1}\ \rm Mpc$ per side (\citealt{2014HIRSCHMANN}, \citealt{2017RAGAGNIN}).
        
        \item At $z\sim 4$ simulations correctly reproduce the main sequence normalisation, but fail to reproduce the starburst population. Indeed, simulations under-predict the SFR of highly star forming protocluster regions by a factor of 4.
        
        \item In our simulations, the normalisation of the main sequence strongly depends on the gas fraction. Comparison with observations suggests that simulations under-predict the gas fraction in galaxies at the peak of the cosmic star formation rate density and consequently the normalisation of the main sequence.
        
        \item In numerical simulations the position on the main sequence depends on both the gas fraction and the star formation efficiency. However, observations of galaxy properties in dense environment are affected by uncertainties on the assumptions needed to derive gas related quantities. Therefore, it remains unclear whether simulations under reproduce starburst galaxy population because of a low gas fraction or a low star formation efficiency.
        
        %not in agreement among each other and strongly depend on the assumptions needed to derive the molecular gas mass. Therefore, it is not clear whether simulations under reproduce starburst galaxy population because of a low gas fraction or a low star formation efficiency.

        \item Our results indicate that the adopted model of star formation (i.e., \citealt{2003SPRINGEL}) reproduces well the self-regulated evolution of quiescent low-redshift star formation but is not suitable to capture violent events like high-redshift starbursts. We verified that our results are robust and the conclusions hold for a wide range of values of the model parameters and do not depend on the implementation of the AGN feedback.

    \end{itemize}

   Finally, we remark that even though simulations tend to under-reproduce the level of SFR at high redshift, the stellar mass at $z=0$ is even higher than what observations suggest (see Fig.~\ref{fig:bcgm}). Therefore, as already pointed out by \cite{2015GRANATO}, the star formation history of protoclusters must be characterised by peaks that are higher and shorter in comparison to numerical simulations. Given the results obtained in this work, it seems unfeasible to achieve this goal without any substantial modification in the model of star formation, as imposing a self-regulated regime of star formation does not allow to reach high enough values of SFRs. The large amount of data that are becoming available at high redshift from instruments like ALMA will help to put constraints on high redshift galaxy properties and to accordingly improve the degree of realism of star formation models implemented in cosmological simulations of galaxy formation. 
   
%================================================================
%================================================================
\begin{acknowledgements}
%================================================================
%================================================================
    We thank the anonymous referee for the careful and constructive reading of the paper and for his/her useful suggestions.
    We thank L. Boco, L. Pantoni, and M. Valentini for helpful discussions.
    We would like to thank Volker Springel for making the GADGET-3 code available to us. We thank Romeel Dav\'e for sharing data from Simba simulations; Quan Guo for sharing data from EAGLE simulation and GALFORM and L-GALAXIES semi-analytical models; Gabriella De Lucia and Fabio Fontanot for sharing data from GAEA semi-analytical model. 
    VB acknowledges support by the DFG project nr. 415510302. This project has received funding from: ExaNeSt and Euro Exa projects, funded by the European Union Horizon 2020 research and innovation program under grant agreement No 671553 and No 754337, the agreement ASI-INAF n.2017-14-H.0; the Consejo Nacional de Investigaciones Cient\'ificas y T\'ecnicas de la Rep\'ublica Argentina (CONICET); the Secretar\'ia de Ciencia y T\'ecnica de la Universidad Nacional de C\'ordoba - Argentina (SeCyT); the European Union Horizon 2020 Research and Innovation Programme under the Marie Sklodowska-Curie grant agreement No 734374, PRIN-MIUR 2015W7KAWC, the INFN INDARK grant. NRN acknowledges financial support from the "One hundred top talent program of Sun Yat-sen University" gr.ant N. 71000-18841229.
    Simulations have been carried out using MENDIETA Cluster from CCAD-UNC, which is part of SNCAD-MinCyT (Argentina); MARCONI at CINECA (Italy), with CPU time assigned through grants ISCRA B, and through INAF-CINECA and University of Trieste - CINECA agreements; at the Tianhe-2 platform of the Guangzhou Supercomputer Center by the support from the National Key Program for Science and Technology Research and Development (2017YFB0203300). The post-processing has been performed using the PICO HPC cluster at CINECA through our expression of interest. 
\end{acknowledgements}
    
%--------------------------------------------------------------------
\bibliographystyle{aa}
\bibliography{biblio}

\end{document}